\documentclass[11pt,a4paper]{article}
\pdfoutput=1
\usepackage{jheppub}
\usepackage{amsmath}
\usepackage{empheq}
\usepackage{float}

\newcommand{\alp}{\phi}

\newcommand{\Neff}{N_\mathrm{eff}}
\newcommand{\oNeff}{\overline{N}_\mathrm{eff}}
\newcommand{\eqsp}{\,}

\preprint{DESY 20-003}

\graphicspath{ {./plots/} }

\title{Robust cosmological constraints on axion-like particles}

\author{Paul Frederik Depta,}
\author{Marco Hufnagel,}
\author{and Kai Schmidt-Hoberg}

\affiliation{DESY, Notkestra\ss e 85, D-22607 Hamburg, Germany}

\emailAdd{frederik.depta@desy.de}
\emailAdd{marco.hufnagel@desy.de}
\emailAdd{kai.schmidt-hoberg@desy.de}

\abstract{Axion-like particles with masses in the keV-GeV range have a profound impact on the cosmological evolution of our Universe, in particular on the abundance of light elements produced during Big Bang Nucleosynthesis. The resulting limits are complementary to searches in the laboratory and provide valuable additional information regarding the validity of a given point in parameter space. A potential drawback is that altering the cosmological history may potentially weaken or even fully invalidate these bounds. 
The main objective of this article is therefore to evaluate the robustness of cosmological constraints on axion-like particles in the keV-GeV region, allowing for various additional effects which may weaken the bounds of the standard scenario. 
Employing the latest determinations of the primordial abundances as well as information from the cosmic microwave background we find that while bounds can indeed be weakened, very relevant robust constraints remain.}

\begin{document}
\maketitle

\section{Introduction}
Axions or axion-like particles (ALPs) naturally arise in many theories beyond the Standard Model (SM) as Pseudo-Goldstone bosons of some broken global symmetry.
In addition to the well-known QCD axion \cite{Peccei:1977hh,Wilczek:1977pj,Weinberg:1977ma,Hook:2014cda,Fukuda:2015ana} ALPs may naturally arise in string compactifications~\cite{Arvanitaki:2009fg,Cicoli:2012sz},
upon breaking of a continuous $R$-symmetry~\cite{Bellazzini:2017neg}, or in the context of the relaxion mechanism~\cite{Graham:2015cka,Espinosa:2015eda}. 
More generally light pseudoscalar particles are very interesting from a phenomenological point of view and a vital ingredient in any BSM model builder's tool box. 
Sufficiently light ALPs, including most variants of the QCD axion, are very long-lived and may well constitute the elusive dark matter particles~\cite{Preskill:1982cy,Abbott:1982af,Dine:1982ah,Arias:2012az} or be responsible for astrophysical anomalies, such as the Universe's unexpected transparency to high-energy $\gamma$-rays~\cite{Meyer:2013pny} or the presence of a mono-energetic X-ray line around 3.5 keV~\cite{Cicoli:2014bfa,Jaeckel:2014qea}, see however~\cite{Hektor:2018lec}.

ALPs with lifetimes much smaller than the age of the Universe -- while evidently no good dark matter candidates --  may nevertheless play an important role in particle physics phenomenology. For example ALPs have been proposed as a possible explanation of the discrepancy between the measured and the theoretically expected anomalous magnetic moment of the muon~\cite{Chang:2000ii,Marciano:2016yhf} or of an apparent resonance observed in particular nuclear transitions of beryllium~\cite{Ellwanger:2016wfe} and more recently in helium~\cite{Krasznahorkay:2019lyl}. 
In addition ALPs may naturally connect thermal DM particles to the SM, i.e.\ act as DM `mediators', while evading strong constraints from direct detection 
experiments~\cite{Nomura:2008ru,Boehm:2014hva,Dolan:2014ska,Hochberg:2018rjs}. It is therefore not surprising that experimental prospects for ALPs have been the subject of many recent studies~\cite{Mimasu:2014nea,Dolan:2014ska,Jaeckel:2015jla,Dobrich:2015jyk,Izaguirre:2016dfi,Knapen:2016moh,Brivio:2017ije,Bauer:2017nlg,Choi:2017gpf,Bauer:2017ris,Dolan:2017osp,Gavela:2019cmq}.

Weakly coupled light particles such as ALPs also have a profound impact on the cosmological evolution of our universe, in particular on the abundance of light elements produced
during Big Bang Nucleosynthesis (BBN)~\cite{Masso:1995tw,Masso:1997ru,Cadamuro:2010cz, Cadamuro:2011fd,Millea:2015qra}. 
The resulting limits on the parameter space of ALPs are complementary to searches in the laboratory and provide valuable additional information regarding the validity
of a given point in parameter space. In the particle physics community, however, cosmological bounds on a given model are often perceived as `soft' in the sense that altering the cosmological history may well weaken or even fully invalidate these bounds. To rectify this perception, the main objective of this article is to evaluate the robustness of cosmological constraints on ALPs in the keV-GeV region, allowing for additional effects which may weaken the bounds of the standard scenario. Here we mainly concentrate on effects which `factorise' from the ALP sector in order to leave the ALP physics unchanged. Specifically we allow for an arbitrary additional relativistic component in the early universe, contributing to $N_\text{eff}$, as well as an arbitrary chemical potential of SM neutrinos. 
We also consider different reheating temperatures $T_\mathrm{R}$ which directly impact the initial ALP abundance. Employing the latest determinations of the primordial 
helium and deuterium abundances~\cite{Tanabashi:2018oca} as well as information from the Planck mission~\cite{Aghanim:2018eyx}  we find that while bounds can indeed be weakened, very relevant robust constraints remain.

This article is organised as follows. In the next section we review the cosmological evolution of ALPs assuming that they are the only relevant degrees of freedom beyond the Standard Model
at temperatures relevant to BBN. In section~\ref{sec:bbn} we will then discuss how we compare cosmological observations with theoretically expected abundances in the ALP-$\Lambda$CDM cosmology, paying particular attention to the various theoretical and experimental uncertainties. In section~\ref{sec:results} we show the resulting bounds on the ALP parameter space, before we
discuss a number of possible modifications to this standard scenario and the overall impact on the limits in section~\ref{sec:extra}. 
We conclude in section~\ref{sec:discussion}. Some technical details regarding the solution of the relevant Boltzmann equations are provided in the appendix.

\section{ALP cosmology}

In this section we briefly review the production and cosmological evolution of ALPs and specify our notations and conventions. We will concentrate on the case in which the 
ALP $\alp$ couples predominantly to photons, but will also comment on more general coupling structures later.  
Following standard conventions the relevant Lagrangian we consider can be written as
\begin{align}
\mathcal{L}_\mathrm{ALP} &= \frac{1}{2} \partial_\mu \alp\partial^\mu \alp - \frac{1}{2} m_\alp^2 \alp^2 - \frac{g_{\alp \gamma}}{4} \alp F_{\mu \nu} \tilde{F}^{\mu \nu}\eqsp,
\end{align}
where $F$ is the field strength of standard electromagnetism and $\tilde{F}$ is its dual. Assuming this is the dominant interaction the lifetime of $\alp$ is simply
\begin{align}
\tau_{\alp \gamma} = \Gamma_{\alp \gamma}^{-1} = \frac{64 \pi}{m_\alp^3 g_{\alp \gamma}^2}\eqsp.
\end{align}

The two most relevant processes connecting $\alp$ to the SM heat bath are Primakoff interactions $q^\pm \alp \leftrightharpoons q^\pm \gamma$ with a charged particle $q^\pm$ 
in the early universe plasma, and ALP (inverse) decays $\alp \leftrightharpoons \gamma \gamma$. It is well known that the photon will acquire a thermal (plasmon) mass $m_\gamma = e T \sqrt{g_{*q} (T)}/6$ with $g_{*q}(T)$ defined via the following sum over all charged fermions
\begin{equation}
\sum_{i \in \{ e, \mu, \dots \}} Q_i^2 n_i(T) = \frac{\zeta(3)}{\pi^2}g_{*q}(T)T^3\eqsp.
\label{eq:gq}
\end{equation}
Here $n_i$ is the number density of particle species $i$ and $Q_i$ is the corresponding electric charge in units of $e$. 
A non-vanishing $m_\gamma$ could potentially inhibit ALP decays, $\alp \to \gamma \gamma$, but we find that this effect is negligible for the interesting region in parameter space and therefore set it to zero in the calculation of the (inverse) decay collision operator $C_\gamma(E, T)$ (see also \cite{Millea:2015qra}).
The Boltzmann equation for the phase-space distribution function $f_\alp(p, t)$ of the ALP can then be written in the form
\begin{align}
\frac{\partial f_\alp(p, t)}{\partial t} - H(t) p \frac{\partial f_\alp(p, t)}{\partial p} = \big[C_q(E, T) + C_\gamma(E, T)\big] \times \big[\bar{f}_\alp(p, T) - f_\alp(p, t)\big]\eqsp\label{eq:BoltzEqPhi}
\end{align}
with H the Hubble rate and $T=T(t)$. The two collision operators $C_q(E, T)$ and $C_\gamma(E, T)$ encode the Primakoff interaction and the ALP (inverse) decay, respectively, and are given by\footnote{The expression for $C_q(E, T)$ is only valid for fermions and does not account for W bosons as well as light hadrons, which is conservative when evaluating bounds on the ALP parameter space as additional interactions would merely prolong the chemical equilibrium of the ALPs and therefore result in stronger constraints for the parameter space of interest.}
\cite{Cadamuro:2010cz, Cadamuro:2011fd}
\begin{align}
C_q(E, T)&\;\;\,\simeq \sum_{i \in \{ e, \mu, \dots \}} \frac{g_{\alp \gamma}^2 \alpha}{16} \ln \left(1 + \frac{[4 E (m_i + 3 T)]^2}{m_\gamma^2 [m_i^2 + (m_i + 3 T)^2]} \right) Q_i^2n_i(T)\eqsp,\\[3mm]
C_\gamma(E, T) &\overset{m_\gamma = 0}{\simeq} \frac{m_\alp}{E \tau_{\alp \gamma}} \left[ 1 + \frac{2 T}{p} \ln \left( \frac{1 - \exp \big[-(E + p)/2 T\big]}{1 - \exp \big[-(E - p)/2 T\big]} \right) \right]\eqsp.
\end{align}

\noindent As the Primakoff interaction depends on the number density of charged particles in the plasma, $n \propto T^3$, it will be most efficient at high temperatures and freezes out
once the interaction rate drops below the Hubble rate, $H \propto T^2$. In contrast, the (inverse) decay rate drops more slowly than the Hubble rate with decreasing temperature and will therefore be effective at late times.

\begin{figure}[t]
	\centering
	\includegraphics[width=0.495\textwidth]{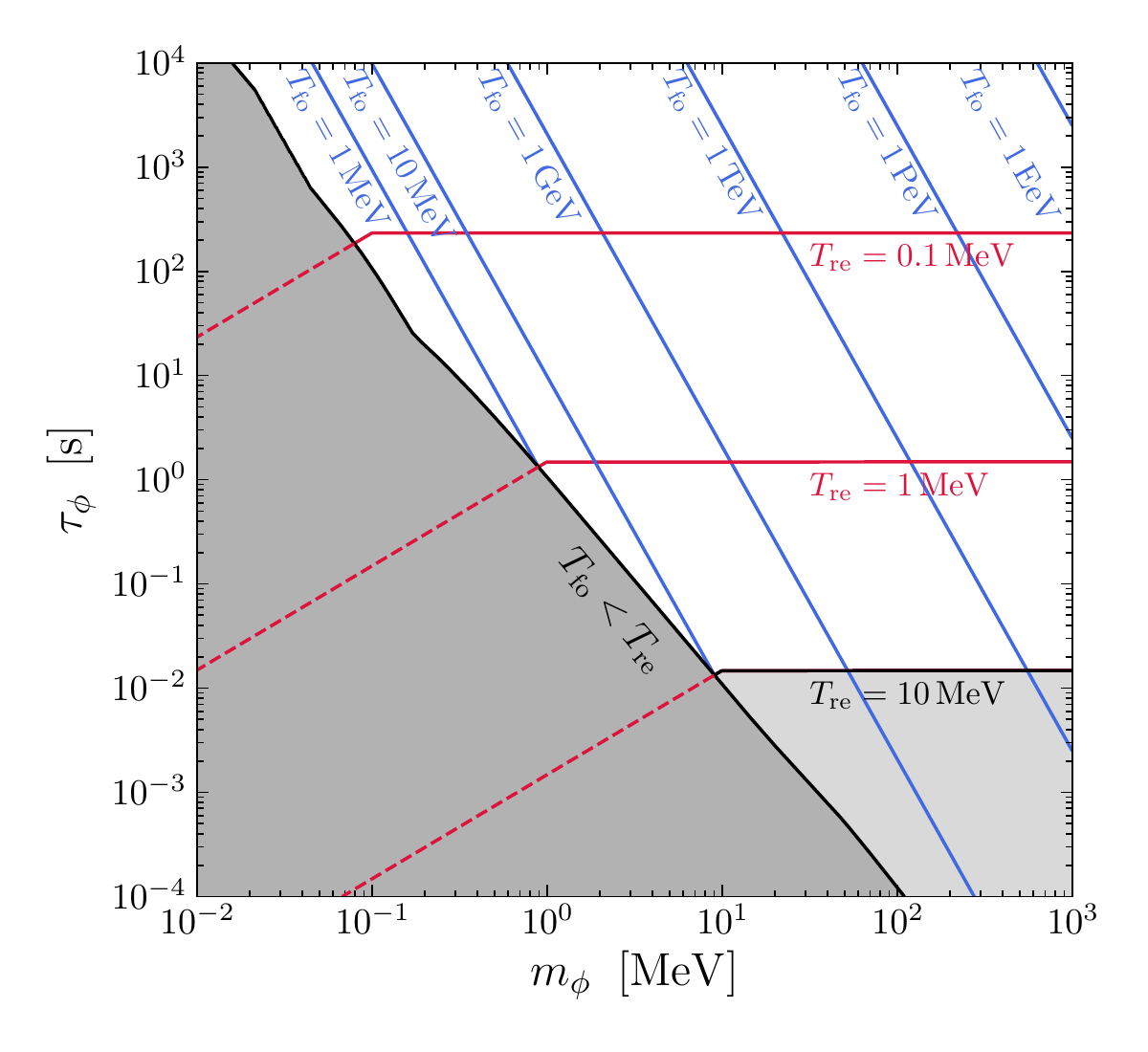}
	\includegraphics[width=0.495\textwidth]{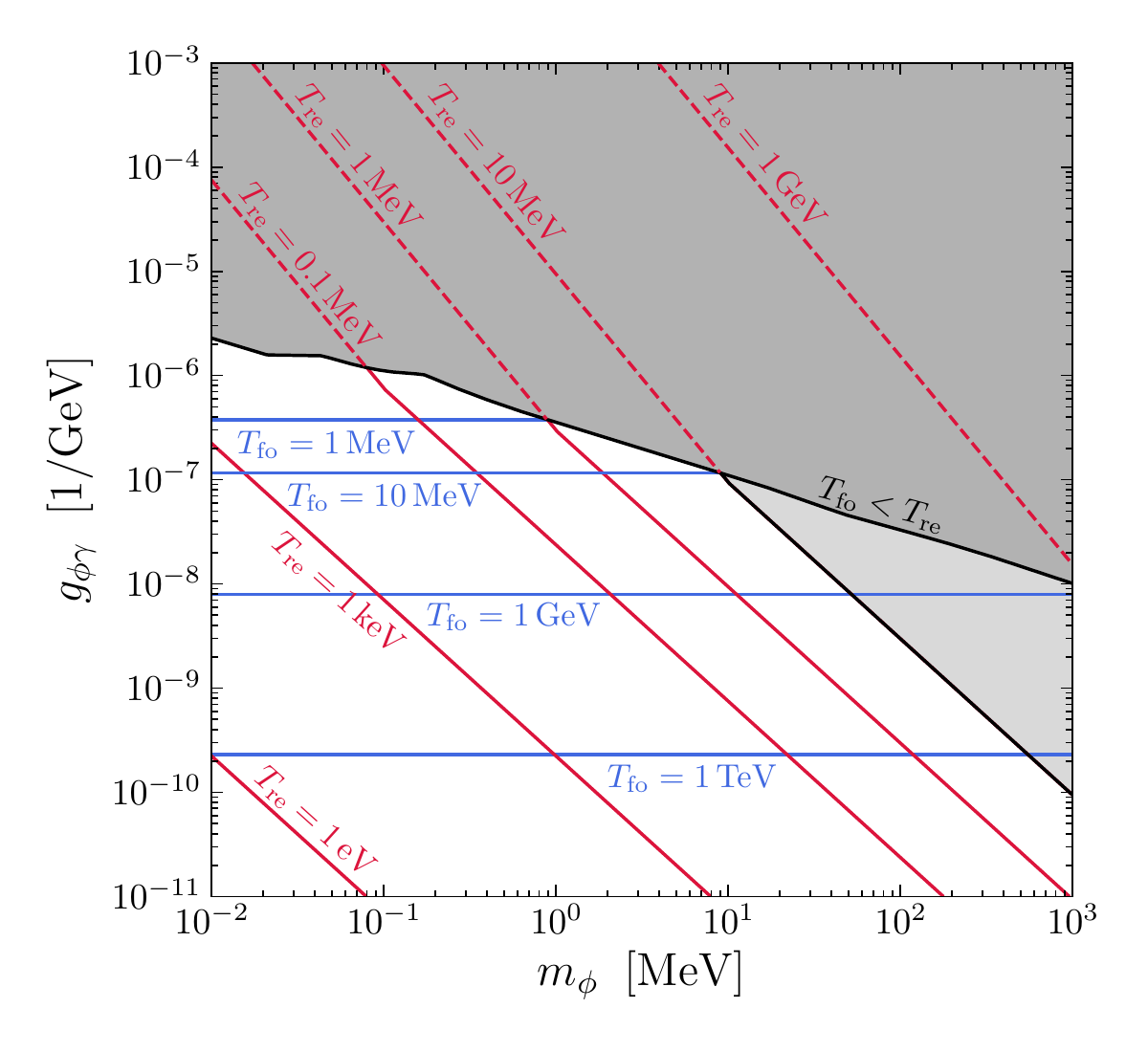}
	\caption{Contours of constant Primakoff freeze-out and (inverse) decay re-equilibration temperatures $T_\mathrm{fo}$ and $T_\mathrm{re}$ as well as $T_\mathrm{fo} = T_\mathrm{re}$ in the $m_\alp-\tau_\alp$ (left) and $m_\alp-g_{\alp \gamma}$ (right) planes.}
	\label{fig:Tfo_Tre}
\end{figure}

For the final results, we solve eq.~\eqref{eq:BoltzEqPhi} numerically without any further approximations (cf.\ appendix~\ref{sec:app}). However, in order to gain a better understanding of the parameter space, it is helpful to derive approximate formulae for both the freeze-out temperature $T_\text{fo}$ as well as for the re-equilibration temperature $T_\text{re}$.
Within 10\% accuracy, the freeze-out temperature can be written as~\cite{Cadamuro:2011fd, Bolz:2000fu}
\begin{align}
T_\mathrm{fo} \simeq 123 \frac{\sqrt{g_{*\rho} (T_\mathrm{fo})}}{g_{*q} (T_\mathrm{fo})} \left( \frac{10^{-9} \mathrm{GeV}^{-1}}{g_{\alp \gamma}} \right)^2 \mathrm{GeV}\eqsp,
\end{align}
where $g_{*\rho}(T)$ is the effective number of SM relativistic degrees of freedom contributing to the energy density.
Analogously, the re-equilibration temperature can be approximated as~\cite{Millea:2015qra}
\begin{align}
T_\mathrm{re} \simeq \begin{cases*} 2.20 \left( \frac{\tau_\alp}{\mathrm{s}} \right)^{-1/2} g_{*\rho} (T_\mathrm{re})^{-1/4} \, \mathrm{MeV} \,, \quad \quad \quad \quad \; \; m_\alp \gtrsim T_\mathrm{re} \\
1.69 \left( \frac{m_\alp}{\mathrm{MeV}} \right)^{1/3} \left( \frac{\tau_\alp}{\mathrm{s}} \right)^{-1/3} g_{*\rho} (T_\mathrm{re})^{-1/6} \, \mathrm{MeV}\,, \ m_\alp \lesssim T_\mathrm{re}\eqsp. \end{cases*}
\end{align}
Qualitatively, the ALP evolution depends on the order of the ALP {\it (i) freezing out from the thermal bath, (ii) becoming non-relativistic} and {\it (iii) re-equilibrating via the (inverse) decay}. 
In figure~\ref{fig:Tfo_Tre} we illustrate how the Primakoff freeze-out temperature $T_\mathrm{fo}$ as well as the (inverse) decay re-equilibration temperature $T_\mathrm{re}$
depend on the mass $m_\alp$ and lifetime $\tau_\alp$ (left) or photon coupling $g_{\alp\gamma}$ (right) of the ALP to roughly map out the different scenarios.
In the region with dark (light) grey the ALP is always (for all temperatures relevant to BBN) in thermal equilibrium, while in the white region the ALP drops out of thermal equilibrium at least for some time before it decays.
To describe the cosmological evolution in more detail one needs to numerically solve eq.~\eqref{eq:BoltzEqPhi}. The technical details of this solution are described in the appendix.

We show the resulting evolution of all relevant energy densities (left) and the corresponding rates (right) for three different representative points in figure~\ref{fig:cosmolator}:
\begin{itemize}
	\item $T_\mathrm{fo} \ll T_\mathrm{re}$: If the Primakoff freeze-out temperature is much smaller than the (inverse) decay re-equilibration temperature (dark grey region in figure~\ref{fig:Tfo_Tre} away from the border), the ALP stays in equilibrium during its entire cosmological evolution, eventually becoming non-relativistic and Boltzmann suppressed. An example point featuring this behaviour is shown in the upper panel in figure~\ref{fig:cosmolator}. As expected the $\alp$ energy density (green line) tracks its equilibrium value (orange dashed line) throughout.
	\begin{figure}[H]
	\centering
	\includegraphics[width=0.495\textwidth]{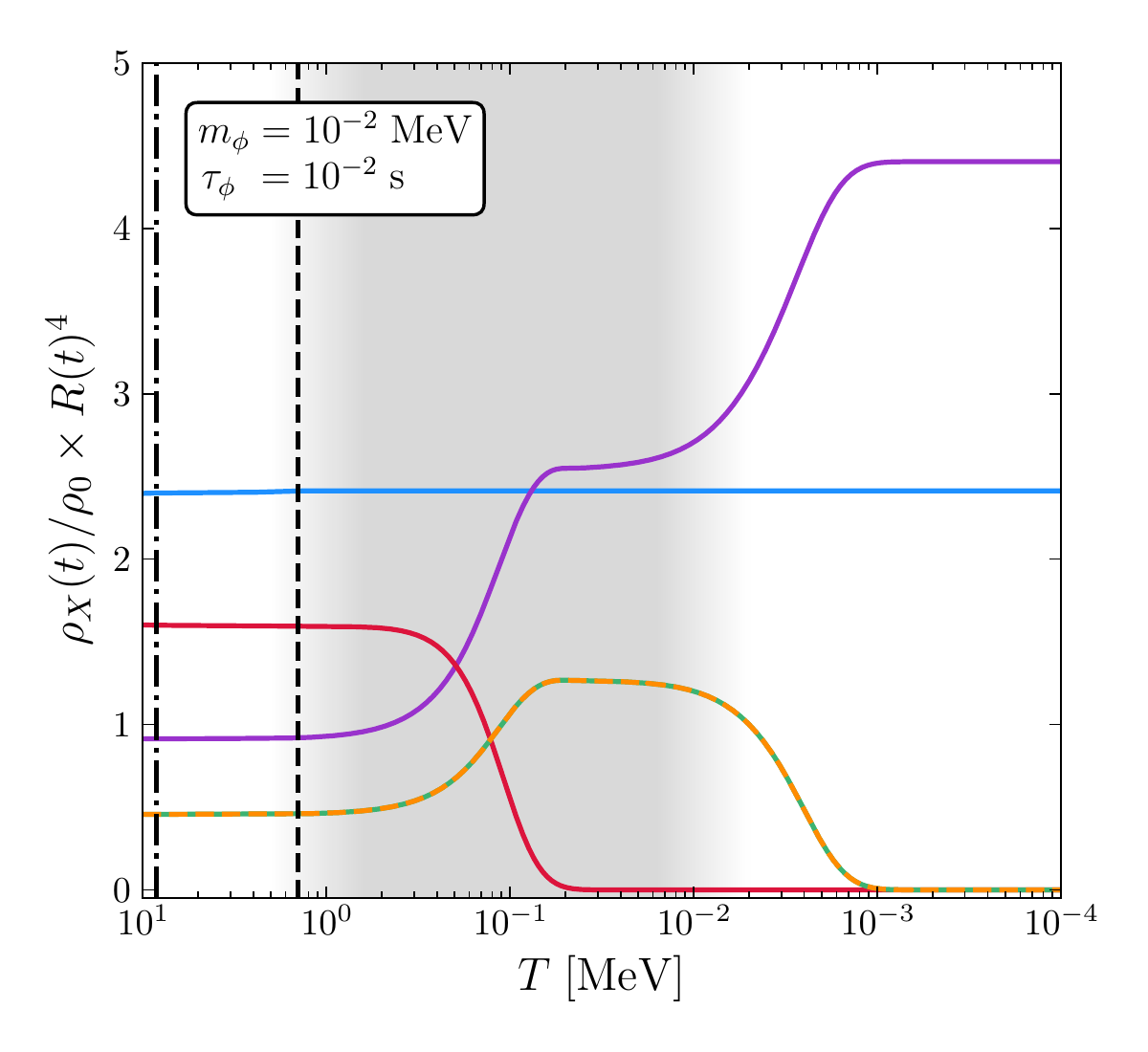}
	\includegraphics[width=0.495\textwidth]{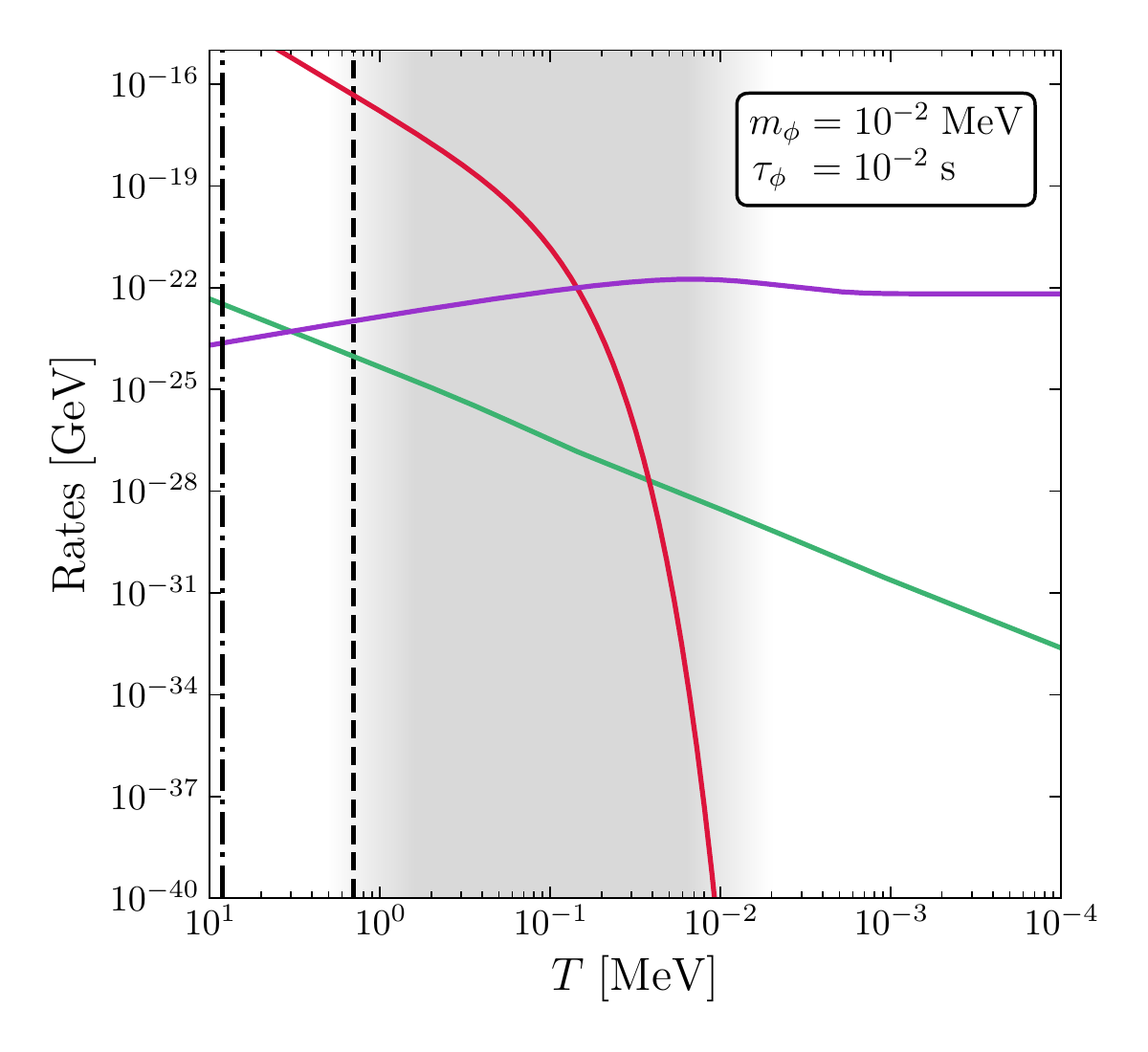}
	\includegraphics[width=0.495\textwidth]{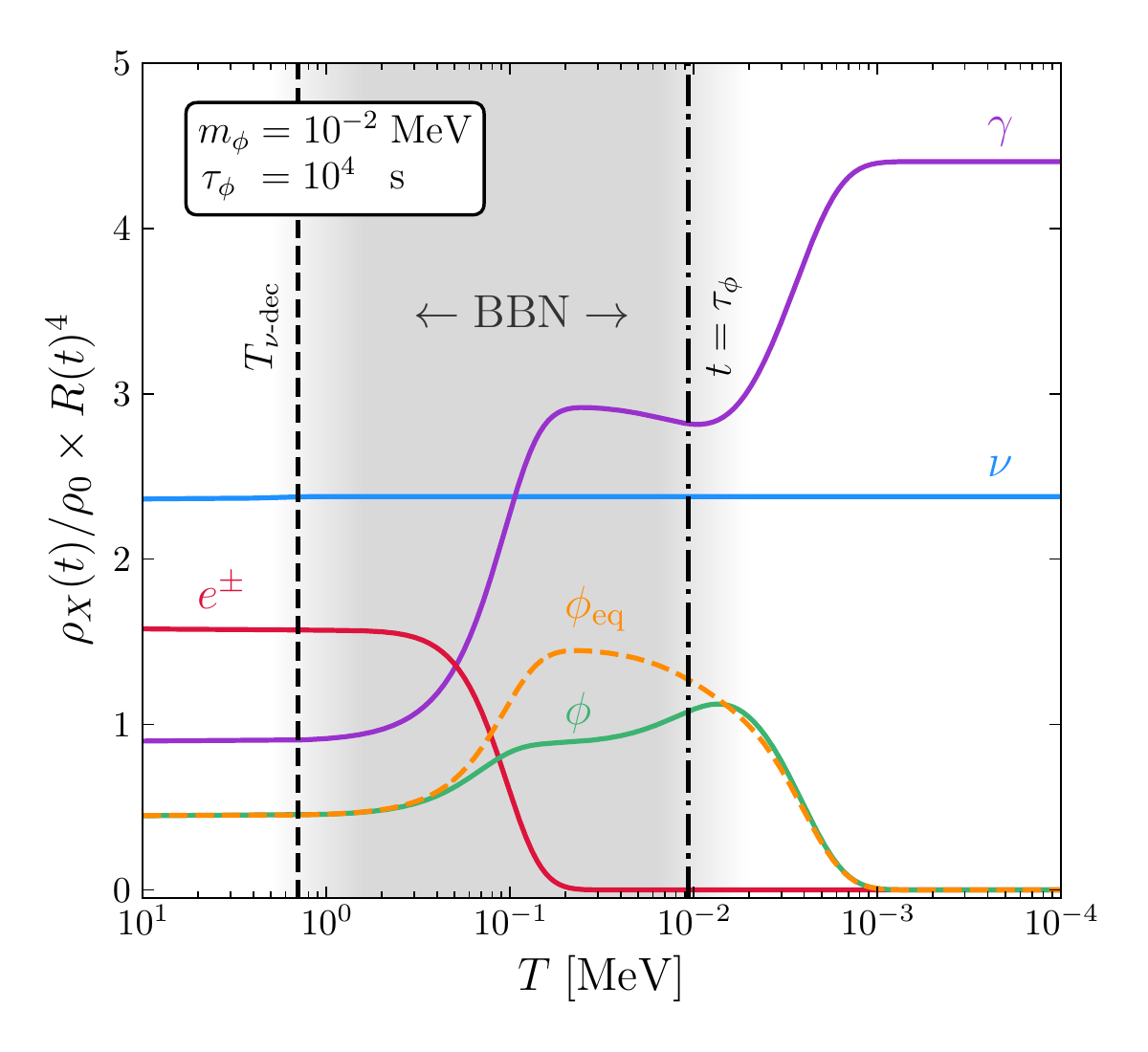}
	\includegraphics[width=0.495\textwidth]{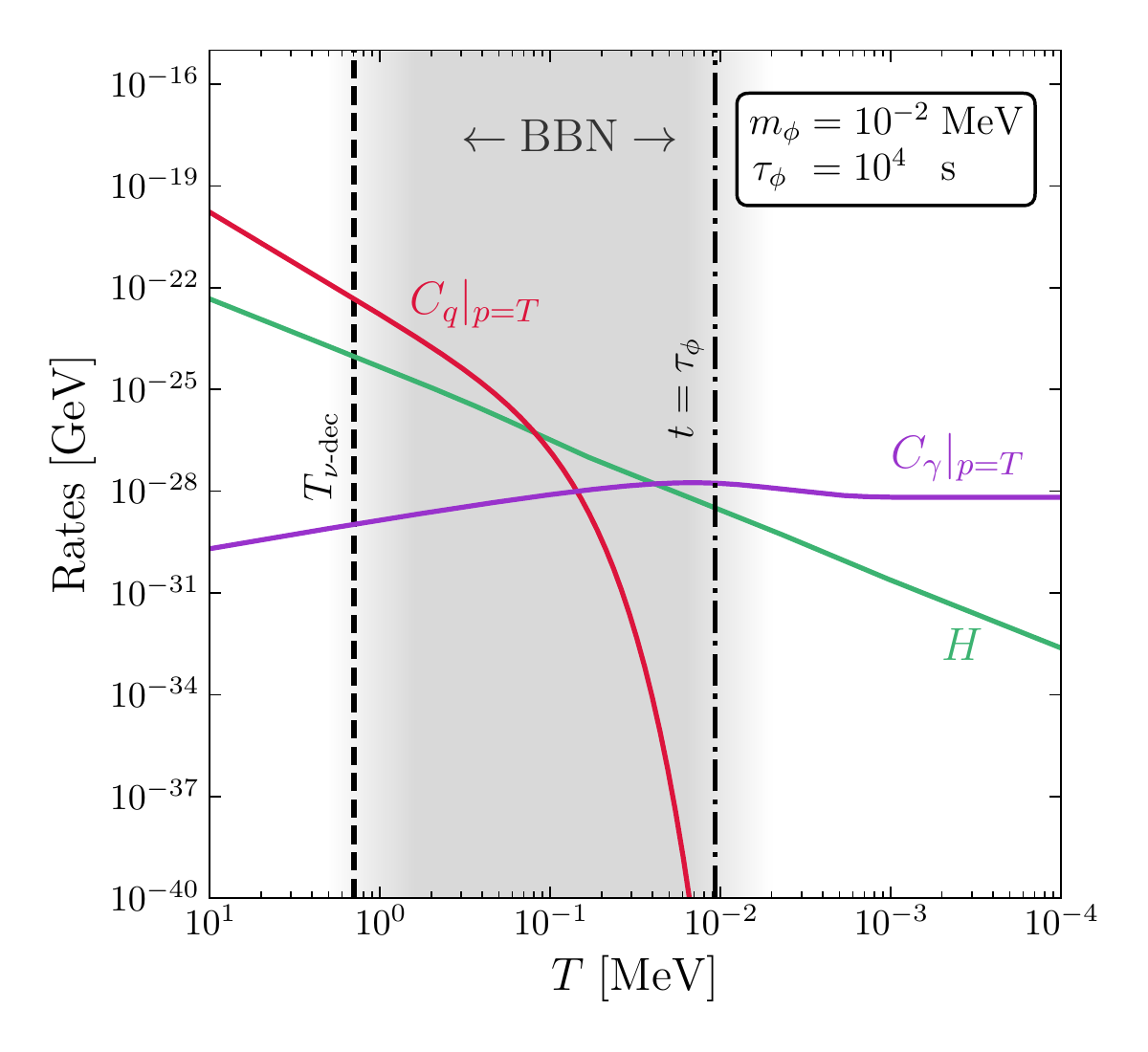}
	\includegraphics[width=0.495\textwidth]{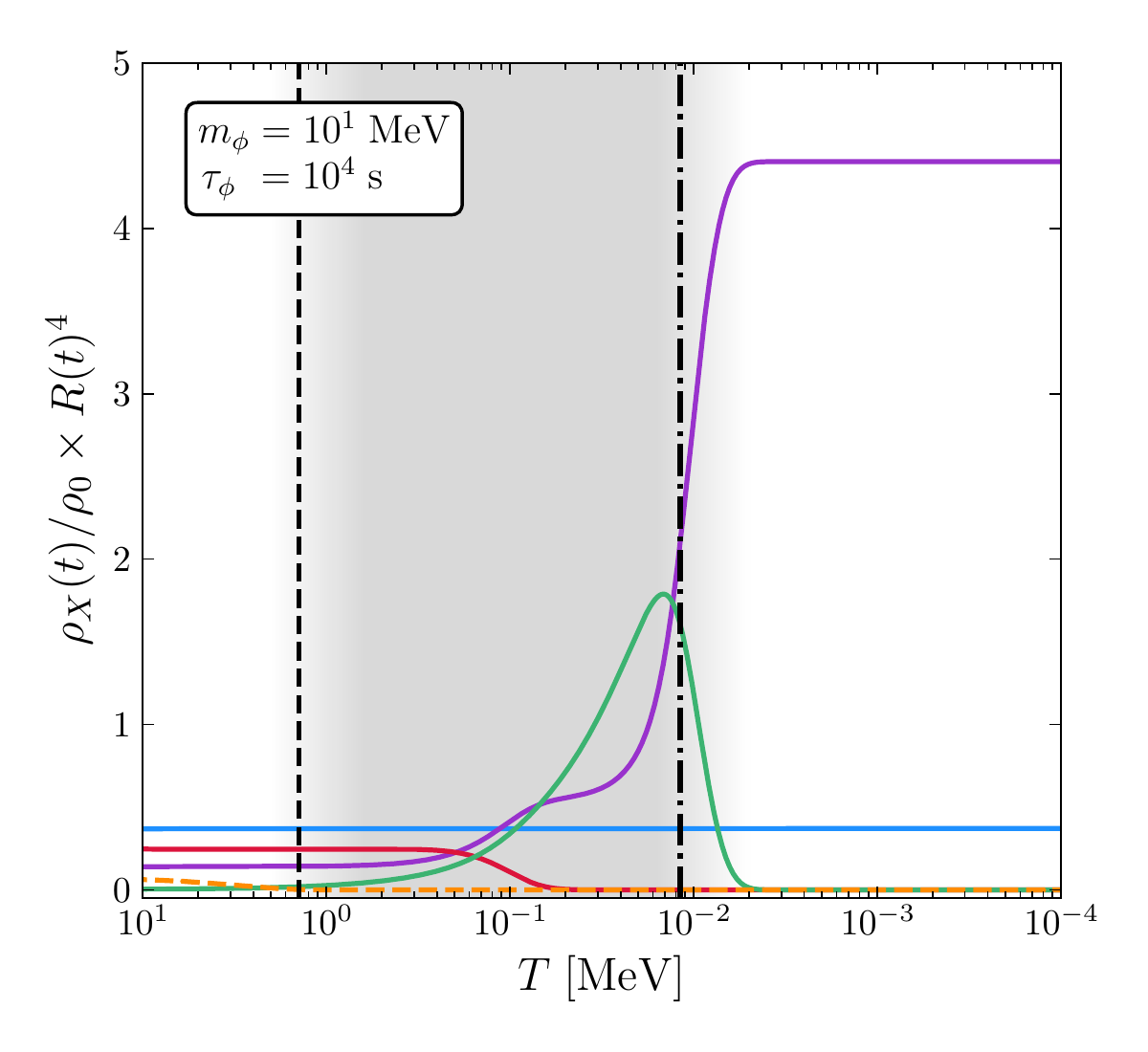}
	\includegraphics[width=0.495\textwidth]{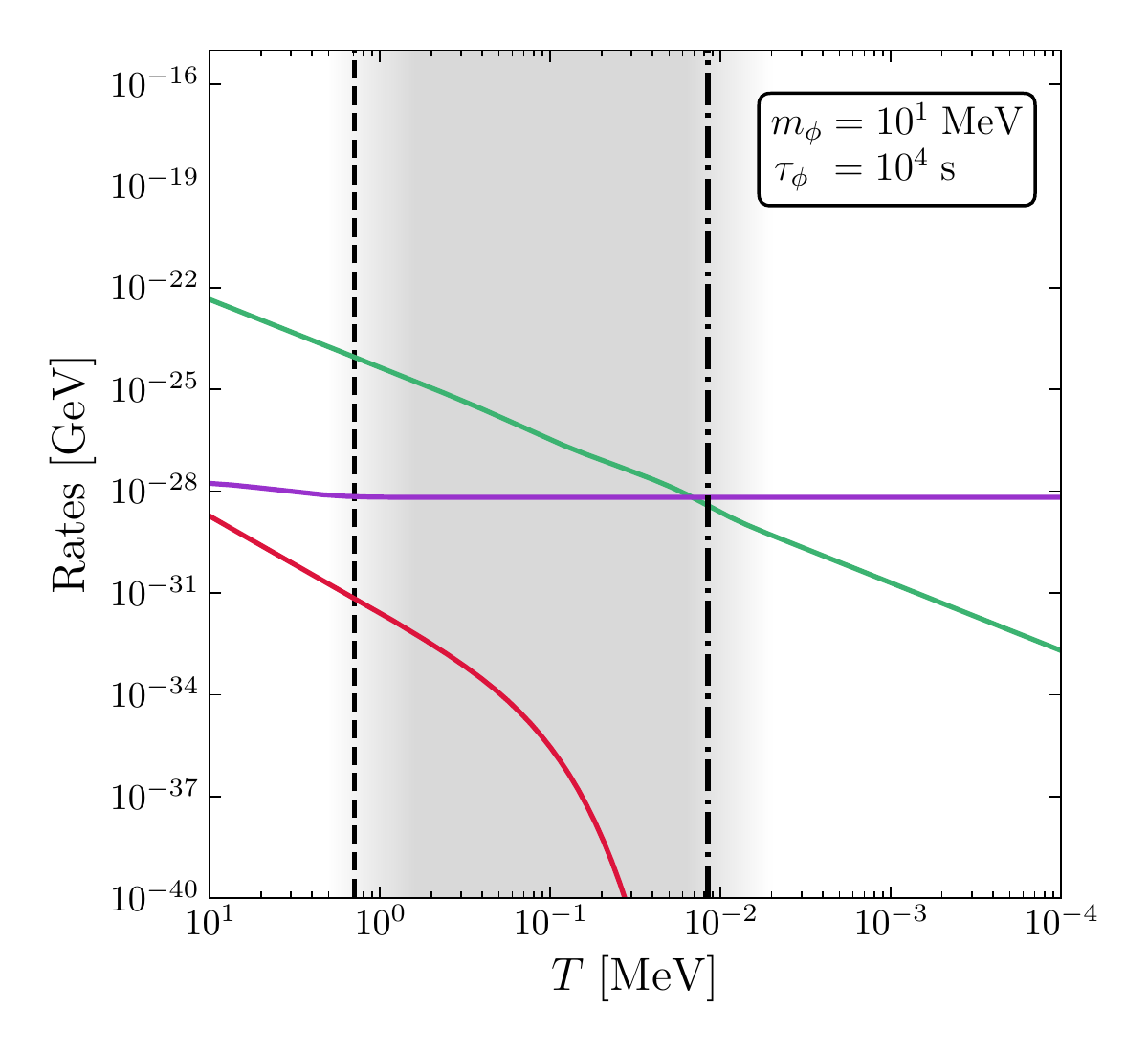}
	\caption{Left: Evolution of the various particle energy densities in the thermal bath for some fixed parameter choices. $\rho_0$ is defined such that $\rho_\nu (t_\mathrm{CMB}) / \rho_0 \times R(t_\mathrm{CMB})^4 = \Neff (t_\mathrm{CMB})$, where $R$ is the scale factor, $\Neff$ is the effective number of neutrinos, and $t_\mathrm{CMB}$ is the time of the cosmic microwave background (CMB). Right: Corresponding rates, where we set $p=T$.}
	\label{fig:cosmolator}
\end{figure}
	\item $T_\mathrm{fo} \sim T_\mathrm{re}$: If the Primakoff freeze-out temperature and the (inverse) decay re-equilibration temperature are similar, different ALP evolutions are possible. One example is displayed in the middle panel in figure~\ref{fig:cosmolator}, where the ALP first decouples from the SM, but still contributes to the total energy density, becomes non-relativistic and decays shortly thereafter.
	\item $T_\mathrm{fo} \gg T_\mathrm{re}$: If the Primakoff freeze-out temperature is much larger than the (inverse) decay re-equilibration temperature (white and light grey region in figure~\ref{fig:Tfo_Tre} away from the border to dark grey region), the ALP decouples from the SM thermal bath around $T_\mathrm{fo}$ and decays around $T_\mathrm{re}$. Note that except for close to the line of $T_\mathrm{fo} = T_\mathrm{re}$ one finds $T_\mathrm{re} < m_\alp$, i.e.\ the ALP becomes non-relativistic before it decays, which leads to an increase of its relative energy density. An example of a parameter point in this region can be found in the lower panel in figure~\ref{fig:cosmolator}.
\end{itemize}

We will now discuss the impact of the ALPs on the predicted abundances of light elements for the standard cosmological evolution before discussing possible alternative cosmological scenarios in section~\ref{sec:extra}.

\section{Primordial element abundances}
\label{sec:bbn}

In the last few years the primordial abundances of light elements have been measured ever more precisely and we use the latest recommendations for the observed abundances of $\mathcal{Y}_\mathrm{p}$ and $\mathrm{D}/{}^1\mathrm{H}$~\cite{Tanabashi:2018oca} as well as for ${}^3\mathrm{He}/\mathrm{D}$~\cite{Geiss2003}\footnote{Note that due to uncertainties in the observation of the primordial abundance of ${}^3\text{He}/{}^1\mathrm{H}$, we only employ the fraction ${}^3\mathrm{He}/\mathrm{D}$ as an upper bound. For a more detailed discussion see~\cite{Depta:2019lbe}.}:
\begin{align}
& \mathcal{Y}_\mathrm{p} \quad & (2.45 \pm 0.03) \times 10^{-1} \label{eq:Yp_abundance} \eqsp,\\ 
& \mathrm{D}/{}^1\mathrm{H} \quad & (2.569 \pm 0.027) \times 10^{-5} \label{eq:D_abundance} \eqsp,\\
& {}^3\mathrm{He}/\mathrm{D} \quad & (8.3 \pm 1.5) \times 10^{-1} \label{eq:3HeH_abundance} \eqsp.
\end{align}
To compare these measurements to the prediction of the ALP cosmology we calculate the expected nuclear abundances with a modified version of \textsc{AlterBBN~v1.4}~\cite{Arbey:2011nf,Arbey:2018zfh}, where the built-in functions for the temperatures $T(t)$ and $T_\nu(t)$ as well as the Hubble rate $H(t)$ are replaced by the cosmological evolution outlined in the appendix. Following the procedure presented in \cite{Hufnagel:2018bjp} we take into account the uncertainties on the nuclear rates in addition to the uncertainties on the measured nuclear abundances to derive the limit on the ALP parameter space.

We also take into account the baryon-to-photon ratio $\eta$ at the time of recombination, and consistently propagate the best-fit value backwards in time using the calculated time-temperature relation. As our model generally predicts a change of the effective number of neutrinos $\Neff$ compared to the SM expectation, it is important to take into account the known correlation between the best-fit values of $\eta$ and $\Neff$ (see also the discussion in~\cite{Millea:2015qra}). 
As the dependence of $\mathrm{D}/\mathrm{^1H}$ on $\eta$ is relatively large one needs to account for the uncertainty in $\mathrm{D}/\mathrm{^1H}$ due to the uncertainty in $\eta$ when confronting the calculated abundances with observations. 

In general, the correlation\footnote{In fact, the results from~\cite{Aghanim:2018eyx} show a correlation between $\Neff$ and $\omega_b = \Omega_b h^2$, which can however be translated to the baryon-to-photon ratio $\eta = 2.7378 \cdot 10^{-8} \Omega_b h^2$. Note that in the following discussion we always refer to $\eta$ at the time of the CMB.} 
between the observed values of $\Neff$ and $\eta$ can be expressed by considering the probability distribution $\mathcal{P}_X$ for the pair $(\eta, \Neff)$
\begin{align}
2 \pi \sigma_\eta & \sigma_{\Neff} \sqrt{1 - r^2} \times \mathcal{P}_X (\eta, \Neff) = \nonumber \\ & \exp \left( - \frac{1}{2 (1 - r^2)} \left[ \frac{(\eta - \overline{\eta})^2}{\sigma_\eta^2} + \frac{(\Neff - \oNeff)^2}{\sigma_{\Neff}^2} -  \frac{2 r (\eta - \overline{\eta}) (\Neff - \oNeff)}{\sigma_\eta \sigma_{\Neff}} \right] \right)\eqsp,
\end{align}
where $\bar{\cdot}$ and $\sigma_\cdot$ denote the mean and variance, while $r$ is the Pearson correlation coefficient. 
Fitting to the $95 \%$ confidence region ellipse (Planck TT,TE,EE+lowE+lensing+BAO) in the $\Omega_b h^2-\Neff$ plane given in figure~26 of~\cite{Aghanim:2018eyx} yields
\begin{align}
\eta = (6.1276 \pm 0.0489 ) \cdot 10^{-10}\eqsp,\quad
\Neff = 2.9913 \pm 0.1690\eqsp,\quad
r = 0.677\eqsp.
\end{align}

A given point in ALP parameter space corresponds to a fixed $\Neff$. Hence, we search for the value of $\eta$ with the optimal (largest) value of $\mathcal{P}_X$, which must be used to fix $\eta$. This gives
\begin{align}
\eta_{\Neff} = \overline{\eta} + r \sigma_\eta \frac{\Neff - \oNeff}{\sigma_{\Neff}}\eqsp.
\end{align}
which we use to fix $\eta$ at the time of the CMB, i.e.\ we set $\eta(t = t_\mathrm{CMB}) = \eta_{\Neff}$. Additionally, we further translate this variation of $\eta$ with $\Neff$ into an uncertainty for the various nuclear abundances. Using linear error propagation for $\mathrm{D/^1H}$ we find\footnote{We used AlterBBN for calculating the value of $\text{d} (\mathrm{D}/\mathrm{^1H}) / \text{d} \eta$.}
\begin{align}
\sigma_{\mathrm{D/^1H}}^\mathrm{eta} = \left| \frac{\text{d}(\mathrm{D}/\mathrm{^1H})}{\text{d} \eta} \sigma_\eta \sqrt{1 - r^2} \right|_{\eta = \eta_{\Neff}} & \approx 0.024 \cdot 10^{-5}\eqsp,
\end{align}
which yields for the total experimental uncertainty
\begin{align}
\sigma_{\mathrm{D/^1H}}^\mathrm{exp} = \sqrt{\left(\sigma_{\mathrm{D/^1H}}^{\mathrm{obs}}\right)^2 + \left(\sigma_{\mathrm{D/^1H}}^\mathrm{eta}\right)^2} \approx 0.036 \cdot 10^{-5}\eqsp.
\end{align}
In principle, we also have to perform this translation for $\mathcal{Y}_\mathrm{p}$ and $\mathrm{^3He/D}$. However, in these cases the error from the variation of $\eta$ is at least one order of magnitude smaller than the observational uncertainty, so we can safely neglect this contribution in all cases of interest.

\section{Results}
\label{sec:results}
\begin{figure}[t]
	\centering
	\includegraphics[width=0.495\textwidth]{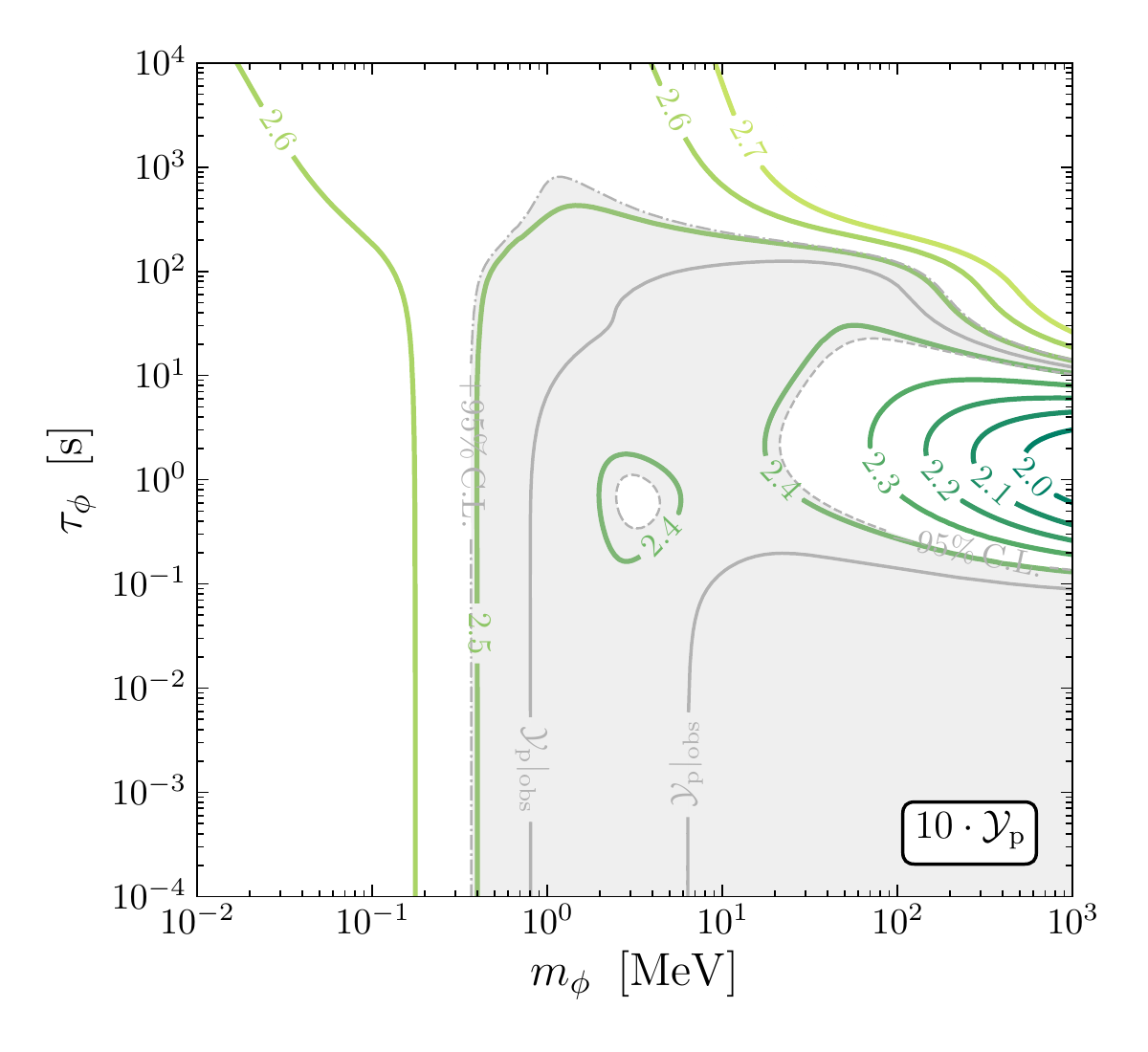}
	\includegraphics[width=0.495\textwidth]{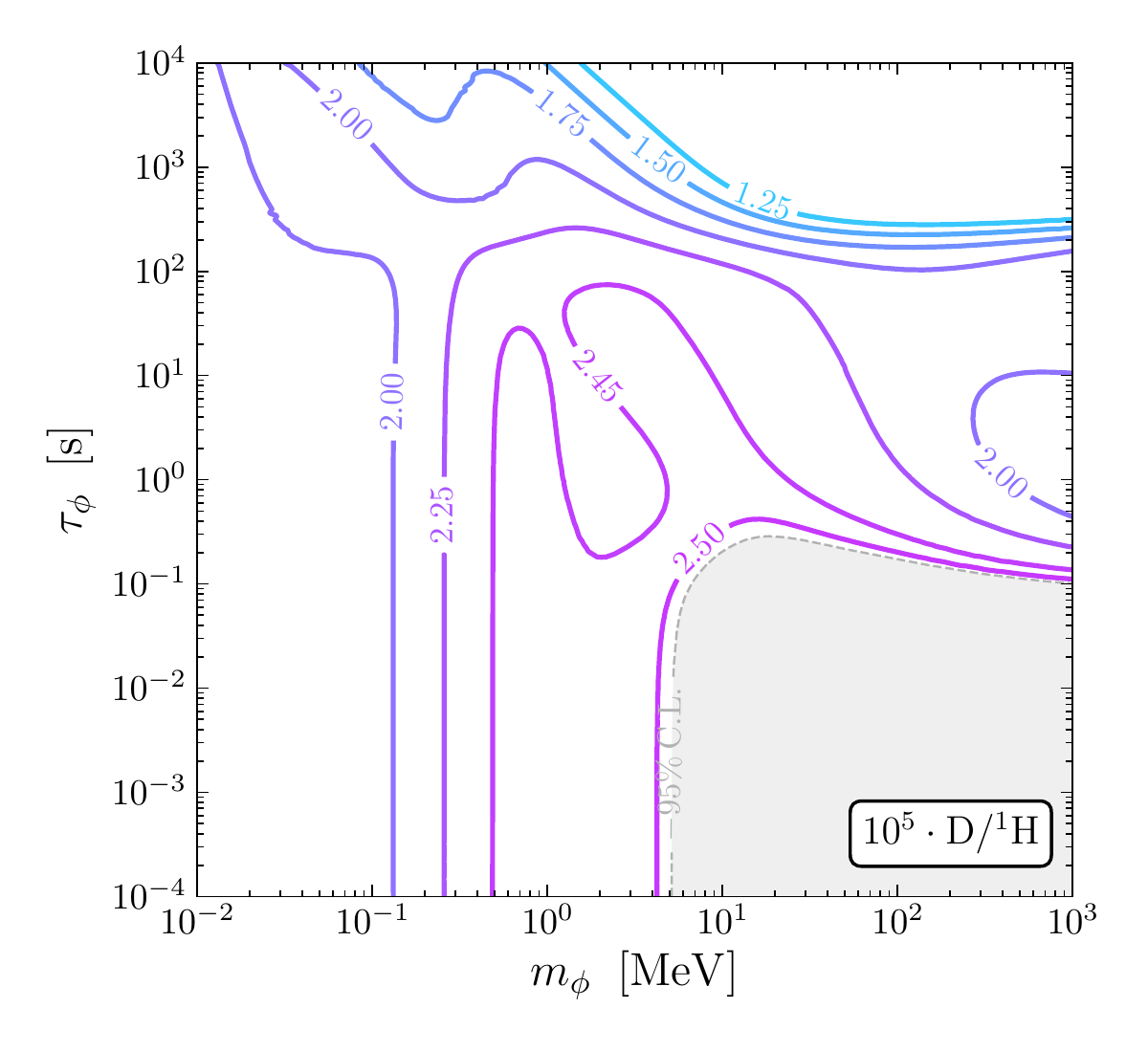}
	\includegraphics[width=0.495\textwidth]{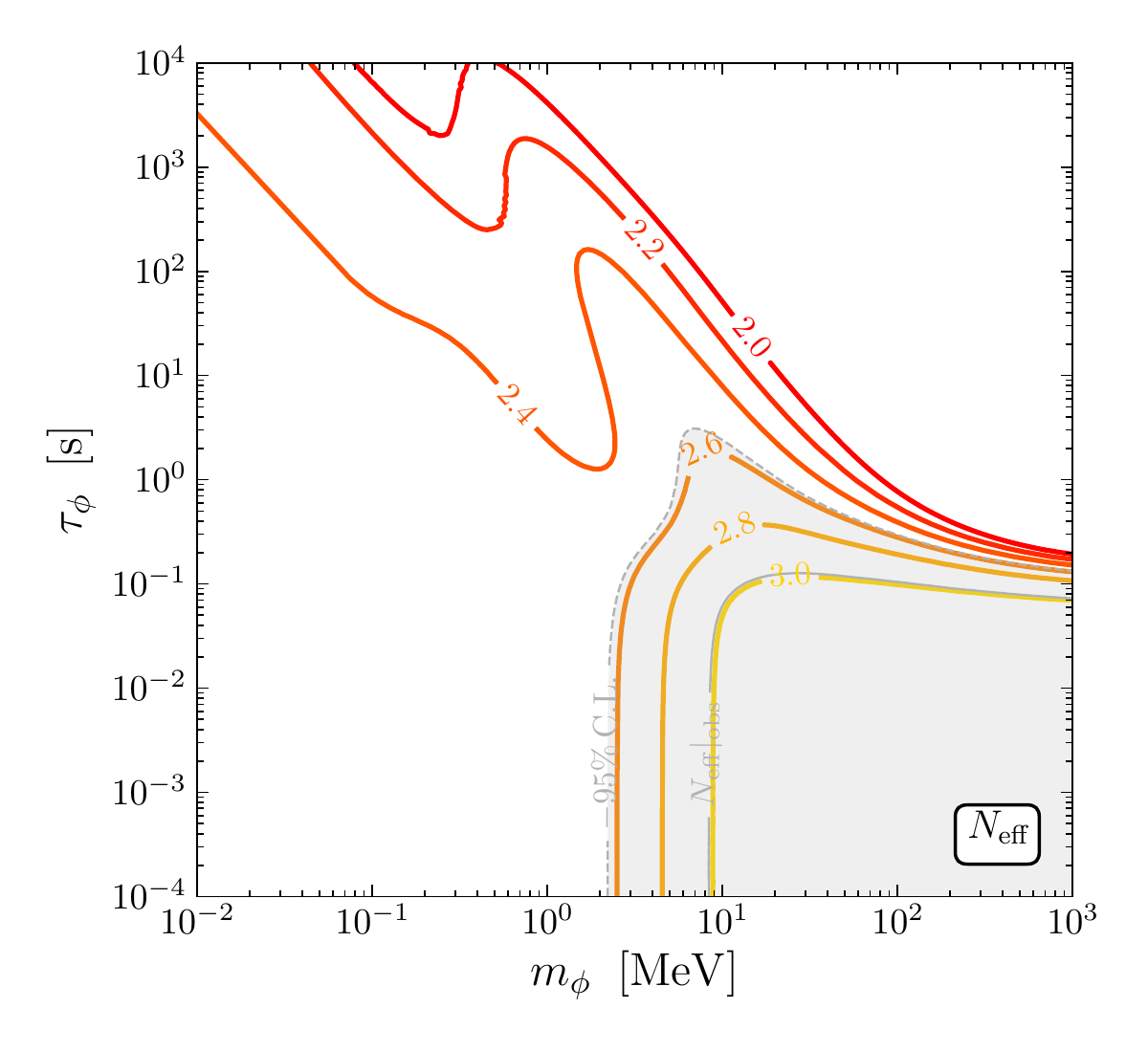}
	\caption{Contours of $10\cdot \mathcal{Y}_\mathrm{p}$ (green), $10^5 \cdot  \mathrm{D}/^{1}\mathrm{H}$ (blue) and $\Neff$ (red) in the $m_\alp - \tau_\alp$ plane. We also indicate where the predicted value agrees with the observed value (grey, solid), as well as where the prediction is $+95\%\,$C.L. (grey, dash-dotted) and $-95\%\,$C.L. (grey, dashed) away from the central value. The grey filled region therefore covers those points in parameter space that are allowed at $2\sigma$, neglecting uncertainties in the nuclear rates.}
	\label{fig:contours}
\end{figure}
To illustrate how the nuclear abundances change as a function of the ALP mass $m_\alp$ and decay time $\tau_\alp$ as well as to connect to previous results in the literature we show the results for  $\mathcal{Y}_\mathrm{p}$ (top left), $\mathrm{D}/^{1}\mathrm{H}$ (top right) and $\Neff$ (bottom) in figure~\ref{fig:contours}.
In particular we indicate the observed values (grey solid lines), as well as their variation at $+95\%\,$C.L. (i.e.\ $95\%\,$C.L. upper limit, grey, dash-dotted) and $-95\%\,$C.L. (i.e.\ $95\%\,$C.L. lower limit, grey, dashed).  Note that the bottom panel shows the limit resulting from the Planck ($\Omega_b h^2-\Neff$) $95\%\,$ confidence region, with the optimal $\eta$ as discussed above. The range of $\Neff$ then corresponds to $\Neff = 2.99 \pm 0.42$ which is somewhat larger than the range corresponding to the central value of $\eta$. 
The grey filled regions therefore na\"ively cover those points in parameter space that are still allowed at $95\%\,$C.L. Note however that the upper panels assume the central value for the nuclear rates. Taking into account the corresponding uncertainties will further enlarge the available allowed parameter space, as can be seen when comparing to figure~\ref{fig:constraints_neff_0} below. Comparing to figure~5 of~\cite{Millea:2015qra}, where also $\mathrm{D}/^{1}\mathrm{H}$, $\mathcal{Y}_\mathrm{p}$, and $\Neff$ are shown, we find larger values for $\mathrm{D}/^{1}\mathrm{H}$, whereas the results agree well for $\mathcal{Y}_\mathrm{p}$ and $\Neff$. This directly translates into a stronger bound resulting from $\mathrm{D}/^{1}\mathrm{H}$ and therefore an overall stronger limit in~\cite{Millea:2015qra} compared to our findings. We will comment below why we believe that our results are correct.

\begin{figure}[t]
	\centering
	\includegraphics[width=0.6\textwidth]{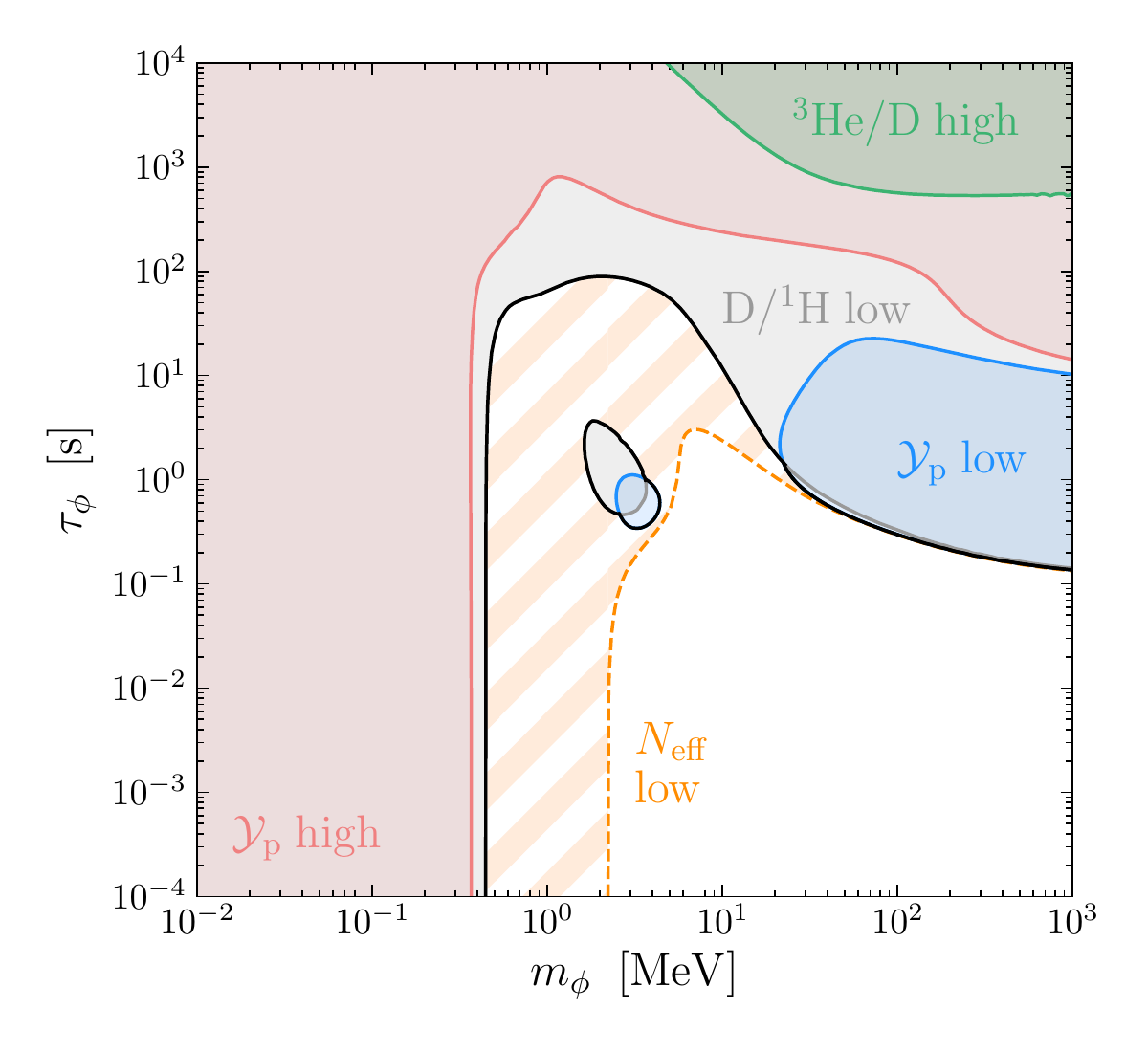}
	\caption{Constraints from BBN and Planck $\Neff$ at $95\%\,$C.L.\ assuming standard cosmological evolution. In addition to the combined limit (solid black line), we also separately list the regions of parameter space which are excluded due to deuterium underproduction (grey), $^{4}$He underproduction (blue), $^{4}$He overproduction (light red), and $^{3}$He overproduction relative to deuterium (green). Values of $\Neff$ that are incompatible with the most recent Planck measurements are indicated by the orange hatched region.}
	\label{fig:constraints_neff_0}
\end{figure}

In figure~\ref{fig:constraints_neff_0} we show the $95\%\,$C.L.\ bounds in the $m_\alp-\tau_\alp$ plane taking into account all relevant uncertainties. We show individual regions which result in $\mathrm{D/{}^1 H}$ underproduction (grey), $\mathcal{Y}_\mathrm{p}$ under-/overproduction (blue/light red), and ${}^3 \mathrm{He/D}$ overproduction (green). We also show the limit resulting from  $\Neff$ as measured by Planck (orange hatched). Note that for the combined limit (black line) we simply employ the envelope of the individual $95\%\,$C.L.\ constraints and do not consider their correlation to calculate a global $95\%\,$C.L.\ limit. This is conservative as the latter procedure would lead to somewhat stronger constraints in regions where two different exclusion lines are close.

For small masses and lifetimes, $\tau_\alp \lesssim 10^{-1}\,\mathrm{s}$ and $m_\alp \lesssim 10\,\mathrm{MeV}$, the ALP remains in thermal equilibrium throughout BBN (cf.\ figure~\ref{fig:Tfo_Tre}).  
We find that in this region, as expected, the resulting constraints agree well with the findings in \cite{Depta:2019lbe}. In particular this implies that our results for $\mathrm{D}/^{1}\mathrm{H}$ reproduces the one in~\cite{Depta:2019lbe} for this limiting case in contrast to the results from~\cite{Millea:2015qra}. We have further cross-checked our results with figure~2 of~\cite{Boehm:2013jpa} and found good agreement. We therefore believe that the $\mathrm{D}/^{1}\mathrm{H}$ constraint in~\cite{Millea:2015qra} is overly stringent due to an overestimation of $\mathrm{D}/^{1}\mathrm{H}$ underproduction. Note that this conclusion also extends to larger lifetimes and masses.
For large masses, $m_\alp \gtrsim 0.1 \, \mathrm{GeV}$, the ALP is no longer in thermal equilibrium and decays after becoming non-relativistic for sufficiently long lifetimes. Unsurprisingly, the smallest ALP lifetime which is constrained in this case approaches a value of $\tau_\alp \sim 0.1 \, \mathrm{s}$ corresponding to the onset of BBN.
Let us finally remark that the inclusion of photodisintegration would not lead to any additional constraints, as a lifetime in excess of $\tau_\alp \gtrsim 6 \times 10^3 \, \mathrm{s}$ (where photodisintegration 
would become relevant, see e.g.~\cite{Hufnagel:2018bjp}) are in any case excluded in this scenario. It may however become relevant if the initial abundance of $\alp$ is sub-thermal, e.g.\ if the reheating temperature after inflation is below the Primakoff freeze-out temperature. These and other possible modifications of the ALP cosmology will be discussed below.

\section{Beyond the vanilla case}
\label{sec:extra}
Until now we have assumed that the cosmological history corresponds to the usual ALP-$\Lambda$CDM cosmology with a high inflationary scale.
In this section we discuss possible deviations from the usually assumed scenario which may weaken the cosmological bounds on axion-like particles.
In particular we will discuss the effect of a low reheating temperature as well as additional contributions to the radiation energy density and non-vanishing neutrino 
chemical potentials. The constraints on the ALP parameter space including these additional effects are shown in  figure~\ref{fig:Treh_multi} and figure~\ref{fig:constraints_neff_opt}.

\subsubsection*{Effects of a low reheating temperature}
\begin{figure}[t]
	\centering
	\includegraphics[width=0.6\textwidth]{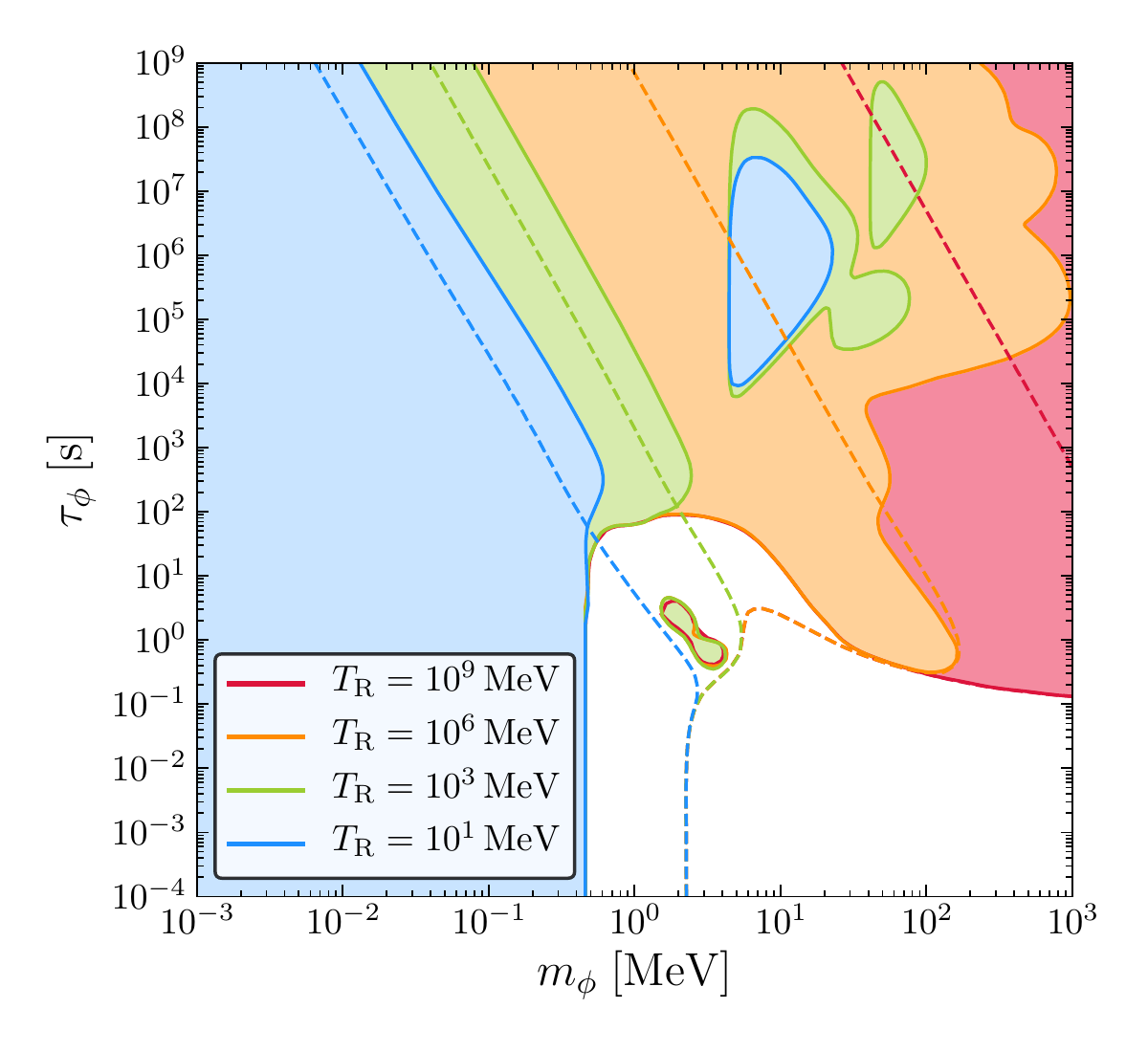}
	\caption{Combined constraint from BBN (full, shaded regions) as well as Planck $\Neff$ (dashed) at $95 \%\,$C.L.\ for different reheating temperatures $T_\mathrm{R}$.}
	\label{fig:Treh_multi}
\end{figure}

Our default calculation of the ALP abundance assumes that the reheating temperature $T_\mathrm{R}$ after cosmic inflation reaches values above the Primakoff freeze-out temperature $T_\mathrm{fo}$, so that ALPs always start out in thermal equilibrium. If this is not the case, ALP production at very large temperatures is suppressed and there is only a `freeze-in' contribution, which scales with $T_\mathrm{R}$ and inversely with $T_\mathrm{fo}$. As there is a very large uncertainty on the reheating temperature, $10 \, \mathrm{MeV} \lesssim T_\mathrm{R} \lesssim 10^{16} \, \mathrm{GeV}$~\cite{Domcke:2015iaa}, it is natural to consider the implication of different $T_\mathrm{R}$ on our limits. We therefore show the combined BBN (full) and Planck $\Neff$ (dashed) constraints at $95 \%\,$C.L.\ for different reheating temperatures in figure~\ref{fig:Treh_multi}, where we conservatively assume that there was no initial ALP abundance at high temperatures (e.g.\ from inflaton decay). For these limits we also 
take into account photodisintegration via the procedure described in~\cite{Hufnagel:2018bjp}, as this is very relevant in constraining even very small ALP abundances.
As we will see below, the BBN constraints are more robust than the Planck $\Neff$ constraints with respect to simple changes of the cosmology, such as an additional contribution to $\Delta \Neff$.

The most pessimistic assumption would be that the reheating temperature does not significantly exceed the minimal value required to be consistent with BBN, which corresponds to the lower end of possible $T_\mathrm{R}$ of about $10 \, \mathrm{MeV}$ (see e.g.~\cite{Hasegawa:2019jsa} for a recent discussion). In the parameter region where the Primakoff interactions freeze out only below $10 \, \mathrm{MeV}$ or where re-equilibration via inverse decays takes place above $10 \, \mathrm{MeV}$, the thermal evolution relevant for BBN is independent of the reheating temperature $T_\mathrm{R}$. This agrees with what we find in figure~\ref{fig:Treh_multi}, where the constraint becomes independent of $T_\mathrm{R}$
for $\tau_\alp \lesssim 10 \, \mathrm{s}$ and $m_\alp \lesssim 0.46 \, \mathrm{MeV}$. 
Overall we see that the constraint for $T_\mathrm{R} = 10 \, \mathrm{MeV}$ (blue region) is significantly weakened compared to the case of high $T_\mathrm{R}$. However, such small reheating temperatures are 
of course an extreme scenario and it may be challenging to successfully have processes required for an overall consistent cosmological picture such as baryogenesis.
Generically one would therefore expect that limits on the ALP parameter space are stronger.

At large lifetimes $\tau_\alp \gtrsim 10^3 \, \mathrm{s}$ and low masses $m_\alp < 4\, \mathrm{MeV}$ the exclusion line approximately scales like the Primakoff freeze-out temperature $T_\mathrm{fo}$ as the production process is governed by the Primakoff process even for $T_\mathrm{fo} > T_\mathrm{R}$. The produced ALP `freeze-in' abundance scales with $T_\mathrm{R}$ and inversely with $T_\mathrm{fo}$. 
Once photodisintegration becomes possible,\footnote{These regions start at twice the threshold energy for the dissociation of nuclei, $m_\alp \approx 2 \times 2.2\, \mathrm{MeV}$ for $\text{D}$ and $m_\alp \approx 2 \times 19.8 \, \mathrm{MeV}$ for ${}^4\text{He}$ and for lifetimes $\tau_\alp \gtrsim 6\times10^3 \, \mathrm{s}$.} even abundances far below the photon density can be constrained~\cite{Poulin:2016anj, Hufnagel:2018bjp}. We therefore find excluded regions detached from the regions excluded by `classic' BBN constraints. For sufficiently high reheating temperatures the constrained regions join as $T_\mathrm{R}$ is large enough for the Primakoff process to produce enough ALPs significantly altering BBN even without considering photodisintegration.
Note that also for high reheating temperatures it is essential to include the effect of photodisintegration, as for $T_\mathrm{fo} \gg T_\mathrm{R}$ the ALP abundance is always very small and the corresponding high-mass region would otherwise be unconstrained.

All in all, we find that constraints indeed become weaker with smaller reheating temperature. Once $T_\mathrm{R} \gtrsim 1 \,\mathrm{PeV}$ the standard scenario is recovered at least until CMB constraints on photon injection become relevant at $\tau_\alp \sim 10^{12}\, \mathrm{s}$. Note that in this discussion we neglected CMB $\mu$- and $y$-distortion constraints relevant for $\tau_\alp \gtrsim 10^6 \, \mathrm{s}$ and $10^8 \, \mathrm{s}$, respectively, which are generically weaker compared to constraints from photodisintegration~\cite{Poulin:2016anj}, albeit not being subject to threshold energies and thus potentially strengthening the bounds for $m_\alp < 4.4 \, \mathrm{MeV}$. 

\subsubsection*{Effects of an additional contribution to $\Neff$}

A rather simple modification of the cosmological scenario would be the addition of extra relativistic degrees of freedom,
which could be completely independent from the ALP sector we consider. As ALP decays happening after neutrino decoupling 
heat up the photon bath compared to the neutrino bath, this results in a {\it reduction} of $\Neff$ compared to the SM expectation and 
one might expect a possible cancellation with an additional positive contribution and hence a weakening of the cosmological bounds. 

To incorporate this effect, we simply add an additional energy density of the form
\begin{equation}
\rho_{\Delta}(T) = 2\Delta \Neff \times \frac{7}{8}\frac{\pi^2}{30} T_\nu(T)^4 \eqsp.
\end{equation}
In terms of the effective number of degrees of freedom, this directly translates to
\begin{align}
g_{*s}^{(\mathrm{SM} + \Delta)}(T) & = g_{*s}^{\mathrm{(SM)}}(T) + 2\Delta \Neff \times \frac78\left( \frac{T_\nu(T)}{T} \right)^3\eqsp, \\
g_{*\rho}^{(\mathrm{SM} + \Delta)}(T) & = g_{*\rho}^{\mathrm{(SM)}}(T) + 2\Delta \Neff \times \frac78 \left( \frac{T_\nu(T)}{T} \right)^4
\end{align}
for the entropy and energy density, respectively. 
We use these modified expressions in the calculation of the time-temperature relation according to the procedure described in~\cite{Hufnagel:2018bjp} (cf.\ also the appendix) to account for an extra contribution to $\Delta \Neff$.
The total value of $\Neff$ at the time of the CMB is then given by
\begin{equation}
\Neff = (3 + \Delta \Neff) \left(\frac{T_\nu(t_\mathrm{CMB})}{T(t_\mathrm{CMB})} \right)^4 \left(\frac{11}{4}\right)^{4/3} \eqsp ,
\end{equation}
where the contributions due to the ALP decay and the additional contribution to $\Delta \Neff$ are combined.

\subsection*{Effects of a non-vanishing neutrino chemical potential}

Another, arguably more contrived, addition to the cosmological picture would be a non-vanishing neutrino chemical potential.
In most scenarios one assumes a vanishing neutrino chemical potential, i.e.\ a vanishing neutrino asymmetry, for BBN~\cite{Iocco:2008va}. This is justified as sphaleron processes, active at temperatures above electroweak symmetry breaking and crucial for many models of baryogenesis, lead to a lepton asymmetry comparable in size to the baryon asymmetry. As neutrinos are relativistic particles for almost all of cosmic history, this implies a negligible neutrino chemical potential. Still, there is no experimental proof of this assumption and there are models for baryogenesis not relying on sphalerons (or alternatively a neutrino asymmetry could have been generated after electroweak symmetry breaking). Thus, even though there is strong theoretical motivation to have a negligible neutrino chemical potential, we treat it as a free parameter and estimate its effect on our constraints.

A non-vanishing neutrino chemical potential $\mu_{\nu_i}$ with  $i \in \{e,\mu,\tau\}$, parametrised by the dimensionless parameters
\begin{equation}
\xi_{\nu_i} \equiv \mu_{\nu_i}(T) / T_{\nu_i}(T)\eqsp,
\end{equation}
which are constant in absence of entropy production in the neutrino sector, modifies standard BBN via two different effects:
On the one hand, there is a change in the effective number of neutrinos, since the corresponding equilibrium distributions change to
\begin{equation}
f_{\nu_i}(E, T) = \frac{1}{\exp\left(E/T_{\nu_i}(T) - \xi_{\nu_i}\right) + 1}\eqsp.
\label{eq:f_nu_xi}
\end{equation}
On the other hand, a non-zero electron neutrino chemical potential influences neutron-proton conversion via the weak interaction, thus enhancing (decreasing) the conversion of neutrons to protons for a positive (negative) $\xi_{\nu_e}$. In equilibrium, the neutron-to-proton ratio is given by
\begin{align}
\frac{n_n(T)}{n_p(T)} = \exp\left(-\frac{m_n-m_p}{T} - \xi_{\nu_e}\right)\eqsp,
\end{align}
implying that at freeze-out, the ratio is suppressed by a factor of $\exp(- \xi_{\nu_e})$ compared to the vanilla case.
To leading order, a smaller neutron density implies smaller values for $\mathcal{Y}_\mathrm{p}$ and $\mathrm{D}/{}^1\mathrm{H}$ as less neutrons are available for the fusion reactions. In the following we will only consider the effect of the electron neutrino chemical potential $\xi_{\nu_e}$, as the effect of the other flavours can always be absorbed in $\Delta \Neff$, which we vary independently at the same time.

\subsection*{Generalised bounds on ALPs}
\begin{figure}[t]
	\centering
	\includegraphics[width=0.495\textwidth]{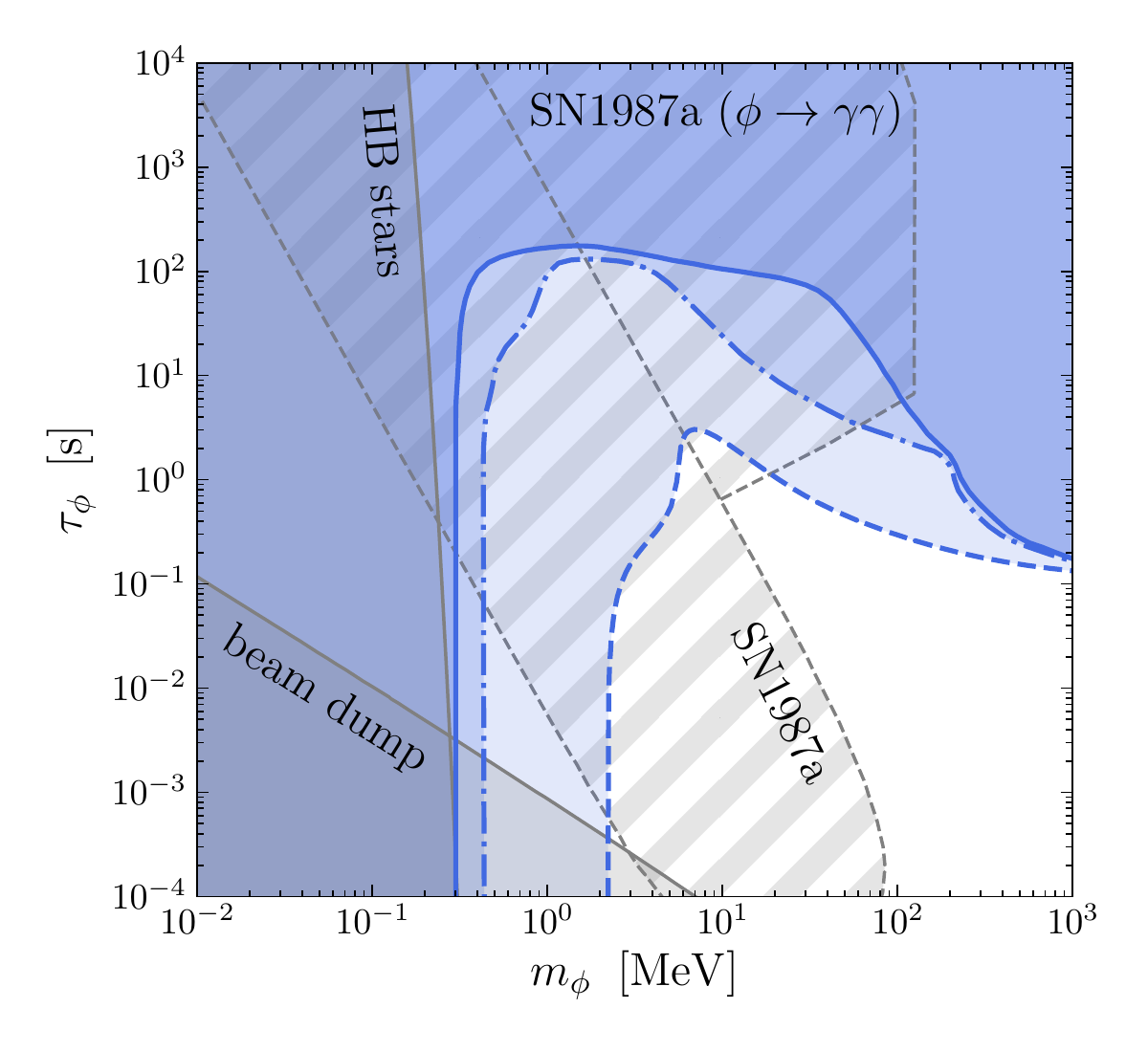}
	\includegraphics[width=0.495\textwidth]{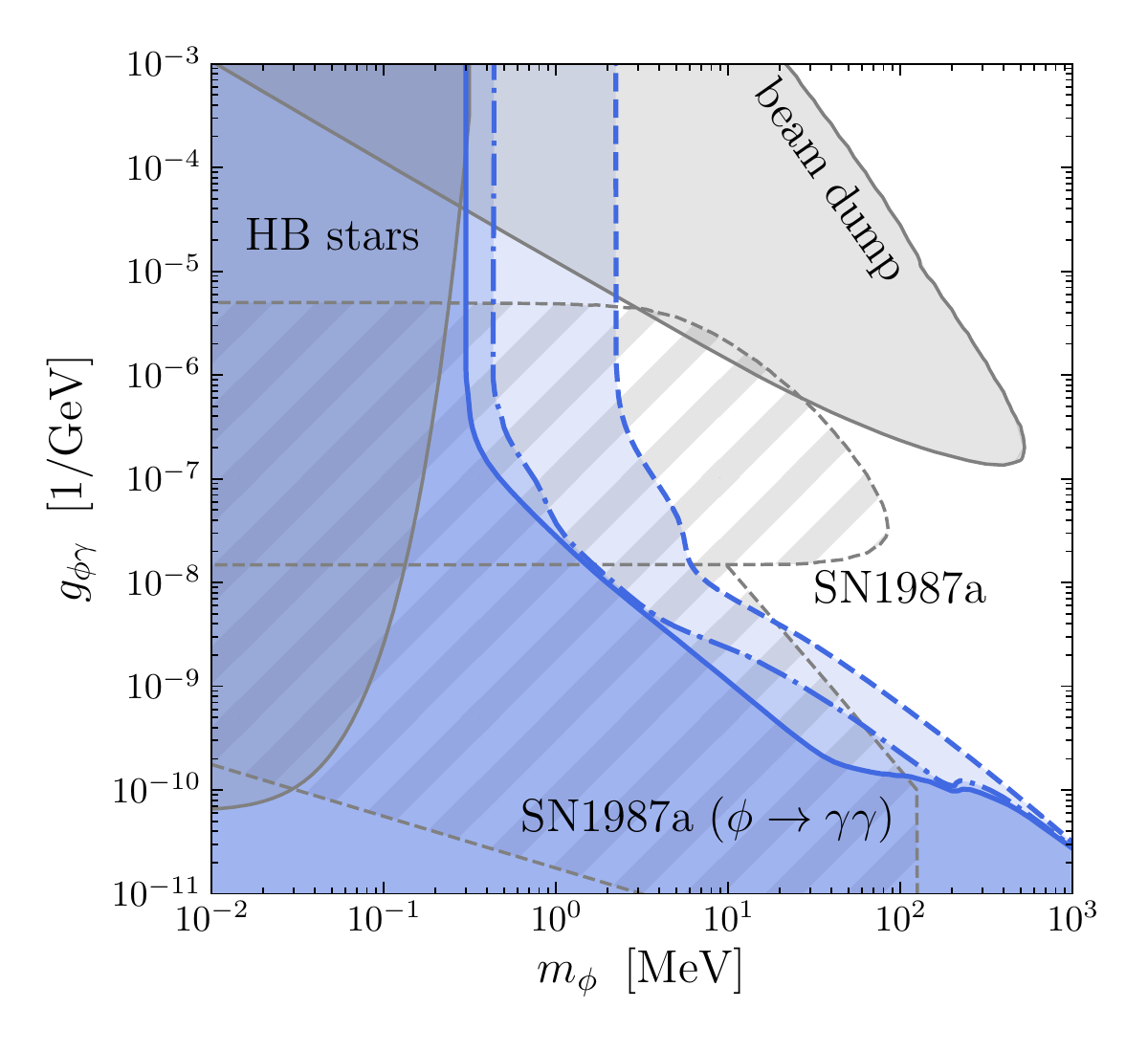}
	\caption{Combined constraints from BBN and Planck $\Neff$ at $95\%\,$C.L. in comparison with complementary constraints from beam dump experiments as well as observations of horizontal branch stars and supernovae. In addition to the case making standard assumptions about the cosmological evolution (blue dashed, cf.\ figure~\ref{fig:constraints_neff_0}), we also show the respective constraints from BBN in case we additionally allow for a variation of $\Delta \Neff$ (blue, dash-dotted) as well as a simultaneous variation of both $\Delta \Neff$ and $\xi_{\nu_e}$ (blue, solid). We follow~\cite{Dobrich:2015jyk} and show the relevant constraints from electron~\cite{Riordan:1987aw, Krasny:1987eb, Dobrich:2017gcm, Bjorken:1988as} and proton beam dumps~\cite{Dolan:2014ska, Bergsma:1985qz, Blumlein:1990ay, Blumlein:1991xh}, SN1987A cooling~\cite{Lee:2018lcj} and visible decays of ALPs produced in SN1987A~\cite{Jaeckel:2017tud} (dashed due to their recently debated reliability~\cite{Bar:2019ifz}), and horizontal branch star cooling~\cite{Cadamuro:2011fd}}
	\label{fig:constraints_neff_opt}
\end{figure}

In figure~\ref{fig:constraints_neff_opt} we show the combined $95\%\,$C.L.\ bound for {\it (i)} the standard ALP scenario (dashed),  {\it (ii)} the ALP scenario with an additional contribution to $\Delta \Neff$
(dash-dotted), and {\it (iii)} the simultaneous addition of $\Delta \Neff$ as well as a neutrino chemical potential $\xi_{\nu_e}$ (full).
Note that for the cases {\it (ii)} and {\it (iii)} we search for a value of $\Delta \Neff$ (and $\xi_{\nu_e}$ if applicable) such that the combined constraint is maximally weakened. As discussed above we only explicitly consider the effect of $\xi_{\nu_e}$ on neutron-proton conversion as the effect of the $\xi_{\nu_i}$ on $\Delta \Neff$ is automatically absorbed by our procedure.

The case {\it (ii)} clearly leads to a weakening of the limit compared to  {\it (i)}. As anticipated, the constraint from Planck on the total $\Neff$ can be partially circumvented. In fact, a value for the total $\Neff$ in agreement with the CMB inferred value is trivially possible, but this does not necessarily lead to the weakest overall bound as there will be simultaneous changes to the BBN predictions. A positive contribution to $\Delta \Neff$ leads to larger abundances $\mathrm{D/{}^1 H}$ and $\mathcal{Y}_\mathrm{p}$. For $0.2 \, \mathrm{GeV} \gtrsim m_\alp \gtrsim 3 \, \mathrm{MeV}$ the combination of these two effects results in the combined constraint being due simultaneously to $\Neff$ being too small and $\mathcal{Y}_\mathrm{p}$ overproduction. For $m_\alp \gtrsim 0.2 \, \mathrm{GeV}$ one finds, depending on the mass, $\mathcal{Y}_\mathrm{p}$ over- or underproduction with $\mathrm{D/{}^1 H}$ underproduction, while still being at the lower boundary for $\Neff$. The required additional $\Delta \Neff$ at the exclusion line ranges from $0.16  \lesssim \Delta \Neff  \lesssim 0.67$ for $m_\alp \leq 10 \, \mathrm{MeV}$ to $0.67   \lesssim \Delta \Neff  \lesssim 22$ for $m_\alp \geq 10 \, \mathrm{MeV}$. In particular the very large values for $m_\alp \geq 10 \, \mathrm{MeV}$ suggest a large possible cancellation, which evidently corresponds to a significant tuning of parameters.

When allowing for a non-vanishing $\xi_{\nu_e}$ in addition to $\Delta \Neff$ (blue full line), the BBN constraints can be further weakened as can be seen in figure~\ref{fig:constraints_neff_opt}.
However, this requires the simultaneous optimisation of a priori independent quantities and therefore to an even more severe tuning in model parameter space. 
Specifically the required values for $\xi_{\nu_e}$ and $\Delta \Neff$ range between $0 \lesssim \xi_{\nu_e} \lesssim 0.12$ and $0.54 \lesssim \Delta \Neff  \lesssim 3.8$ for $m_\alp \lesssim 10 \, \mathrm{MeV}$  and  between $0.12 \lesssim \xi_{\nu_e} \lesssim 0.35$ and $3.8 \lesssim \Delta \Neff  \lesssim 29$ for $m_\alp \gtrsim 10 \, \mathrm{MeV}$.

\subsection*{Comment on more general ALP coupling structures}
Let us finally comment on more general ALP coupling structures which may be naturally present, depending on the UV completion of the ALP model. In this work we concentrated on the coupling of $\alp$ to two photons, $g_{\alp \gamma}$. In more general coupling scenarios, the ALP might be coupled to the gluon field strength in a similar way as to photons and furthermore with a derivative coupling to an axial vector fermion current. The impact of these couplings on cosmological constraints on ALPs has previously been discussed in~\cite{Cadamuro:2011fd}. 
In general the limits are expected to be very similar to the ones we have discussed above if one interprets the ALP lifetime $\tau_\alp$ as the total lifetime. In fact, for the parameter regions 
in which the ALP is always in equilibrium, corresponding to the grey region in figure~\ref{fig:Tfo_Tre}, the limits will directly apply. 
In the rest of parameter space the limits can become weaker or stronger, depending on which additional effects dominate. Specifically, additional couplings in general imply
\begin{itemize}
\item larger production rates due to additional production channels (barring an interference e.g.\ between Primakoff and Compton processes, $q^\pm + \alp \rightleftharpoons q^\pm + \gamma$),
\item ALP freeze-out at smaller temperatures (this can decrease the final energy density if freeze-out happens after the ALP becomes non-relativistic and therefore weaken the constraints),
\item other final states from the ALP decay which will have somewhat different effects on the primordial abundances.
\end{itemize}
For masses $m_\alp < 2 m_\mu$ the only possible two-body final states in addition to photons are electrons and neutrinos. 
However, due to the Yukawa-like coupling structure of $\alp$ to SM fermions the decay into neutrinos is completely negligible and also the decays to electrons
are naturally suppressed, implying that the final state is basically unchanged from what we discussed above.
For heavier ALPs decays into muons and hadrons are kinematically allowed, and while constraints from deuterium and ${}^4 \mathrm{He}$ underproduction may initially weaken as e.g.\ pions will increase the neutron-proton ratio~\cite{Jedamzik:2006xz}, generically hadro-dissociation will lead to significantly stronger bounds on the ALP abundance.

\section{Conclusions}
\label{sec:discussion}

In this article we have discussed cosmological constraints on axion-like particles taking into account the most recent observations of primordial abundances as well as results from
the Planck satellite, where we have paid special attention to the involved theoretical and experimental uncertainties. In particular we have addressed the question how much a changed
cosmological history could weaken these limits, where we concentrated on effects which factorise from the ALP sector in order to leave the associated physics unchanged.
Specifically we discussed the effect of 
\begin{itemize} 
\item a low reheating temperature, where in large regions of parameter space the ALP never reaches thermal equilibrium, suppressing the initial ALP abundance,
\item the addition of independent relativistic degrees of freedom contributing to $\Delta \Neff$,
\item a non-vanishing chemical potential of SM neutrinos, $\xi_{\nu_i}$.
\end{itemize}
Our main result are shown in figure~\ref{fig:Treh_multi} and figure~\ref{fig:constraints_neff_opt}. 
It can be seen that for low reheating temperatures viable parameter space opens up towards large ALP masses and long ALP lifetimes.
Assuming a high reheating temperature, limits on axion-like particles can nevertheless be evaded for some parts of parameter space, 
although significant and robust constraints remain, even if additional effects are tuned to allow for a maximal cancellation of different bounds.

\acknowledgments

We would like to thank Valerie Domcke, Camilo Garcia-Cely, Felix Kahlhoefer and Joerg Jaeckel for helpful discussions.
This work is supported by the ERC Starting Grant `NewAve' (638528) as well as by the 
Deutsche Forschungsgemeinschaft under Germany's Excellence Strategy -- EXC 2121 `Quantum Universe' -- 390833306.

\appendix
\section{Solution of the Boltzmann equation}
\label{sec:app}

In order to solve eq.~\eqref{eq:BoltzEqPhi}, we first introduce the variable $p_\star = p \cdot R(t)$, where $R$ is the scale factor normalised to $R(t_0) = 1$ at the time $t_0$ we start the calculation. This gives
\begin{align}
\frac{\mathrm{d} f_\alp(p_\star, t)}{\mathrm{d} t} = \big[C_q(E, T) + C_\gamma(E, T)\big] \times \big[\bar{f}_\alp(p_\star, T) - f_\alp(p_\star, t)\big]\eqsp.
\end{align}
Note that $E = E(p_\star, t) = \sqrt{m_\alp^2 + p_\star^2 / R(t)^2}$ and again $T = T(t)$. This differential equation is solved by
\begin{align}
f_\alp (p_\star, t) = f_\alp (p_\star, t_0) \exp \left( \int_{t_0}^t \mathrm{d} t_1 \alpha (p_\star, t_1) \right) + \int_{t_0}^t \mathrm{d} t_1 \beta (p_\star, t_1) \exp \left( \int_{t_1}^t \mathrm{d} t_2 \alpha (p_\star, t_2) \right)\eqsp, \label{eq:IntSolPhi}
\end{align}
where
\begin{align}
\alpha (p_\star, t) &= - C_q(E, T) - C_\gamma(E, T)\eqsp, \\[3mm]
\beta (p_\star, t) &= \big[C_q(E, T) + C_\gamma(E, T)\big] \times \bar{f}_\alp(p_\star, T)\eqsp.
\end{align}

We numerically solve eq.~\eqref{eq:IntSolPhi} using a Simpson rule after transforming to $\log (t/t_0)$ for the integral over $\alpha$ and a modified trapezoidal rule assuming that the integrand is approximately piecewise linear in $\log-\log$ space (i.e.\ piecewise follows a power-law) for the integral involving $\beta$. The spectrum as a function of physical momentum can be obtained via $f_\alp (p, t) = f_\alp(p_\star / R(t), t)$, which is related to the energy and number densities $\rho_\alp$ and $n_\alp$ via
\begin{align}
\rho_\alp (t) &= \int \frac{\mathrm{d}^3 p}{(2 \pi)^3} E f_\alp(p, t)\eqsp, \quad n_\alp (t) = \int \frac{\mathrm{d}^3 p}{(2 \pi)^3} f_\alp(p, t)\eqsp.
\end{align}

In the standard scenario, where ALPs once were in thermal equilibrium via the Primakoff interaction, we start our calculation at a temperature $T_0 = 20 \, T_\mathrm{fo}$ and corresponding time $t_0$ with
\begin{align}
f_\alp (p_\star, t_0) \equiv \bar{f}_\alp(p_\star, T_0)
\end{align}
in order to properly take into account Primakoff freeze-out starting from the equilibrium distribution. If $20 \, T_\mathrm{fo} < 10 \, \mathrm{MeV}$ we use $T_0 = 10 \, \mathrm{MeV}$ to start the calculation before the onset of BBN.

When considering a specific reheating temperature $T_\mathrm{R}$, we start the calculation at a temperature $T_0 = T_\mathrm{R}$ and corresponding time $t_0$ with
\begin{align}
f_\alp (p_\star, t_0) \equiv 0 \eqsp,
\end{align}
conservatively assuming that no ALPs are produced during reheating. Note that if $T_\mathrm{fo} < T_\mathrm{R}$, the Primakoff interaction will quickly bring the ALPs into equilibrium.

In any case, we follow the procedure of~\cite{Hufnagel:2018bjp} to calculate the evolution of the SM temperature during the ALP decay using comoving energy and entropy conservation
\begin{align}
\dot{\rho}_\alp + 3 H (\rho_\alp + p_\alp) = -\dot{\rho}_\mathrm{SM} - 3 H (\rho_\mathrm{SM} + p_\mathrm{SM})  \eqsp, 
\end{align}
where $\rho_\alp$ ($\rho_\mathrm{SM}$) and $p_\alp$ ($p_\mathrm{SM}$) are the $\alp$ (SM) energy density and pressure and $g_\alp = 1$ is the number of internal degrees of freedom of $\alp$. If an additional contribution to $\Delta \Neff$ and a neutrino chemical potential $\xi_{\nu_e} \equiv \mu_{\nu_e} / T_{\nu_e}$ is considered, we additionally implement them in all relevant quantities to quantify their effect on the cosmological constraints.

\bibliography{refs}
\bibliographystyle{JHEP}

\end{document}